\documentclass[reprint,aps,prd,twocolumn,notitlepage,nofootinbib,showpacs,preprintnumbers,superscriptaddress]{revtex4-1}
\usepackage[utf8]{inputenc}
\usepackage[T1]{fontenc}
\usepackage{bm}
\usepackage{amsmath,amssymb}
\usepackage{esint}
\usepackage{color}
\usepackage{graphicx}% Include figure files
\PassOptionsToPackage{normalem}{ulem}
\usepackage{ulem}

\usepackage{times}
\usepackage{hyperref}
\begin{document}

\title{Coupling between Galileon and massive gravity with composite metrics}

\author{Xian Gao}%
    \email[Email: ]{gao@th.phys.titech.ac.jp}
    \affiliation{%
        Department of Physics, Tokyo Institute of Technology,\\ 
        2-12-1 Ookayama, Meguro, Tokyo 152-8551, Japan}

\author{Daisuke Yoshida}%
\email[Email: ]{yoshida@th.phys.titech.ac.jp}
\affiliation{%
	Department of Physics, Tokyo Institute of Technology,\\ 
	2-12-1 Ookayama, Meguro, Tokyo 152-8551, Japan}

\date{\today}

\begin{abstract}
	We investigate the coupling between a Galileon scalar field and massive gravity through composite metrics.
	We derive the full set of equations of motion for a flat Friedmann-Robertson-Walker background, and study linear perturbations around it.
	Generally, the nonminimal coupling with the composite metric will excite all 6 degrees of freedom of the spatial metric perturbations, one of which may correspond to the Boulware-Deser ghost.
\end{abstract}


\maketitle
%\tableofcontents

% % % % % % % % % %

\section{Introduction}

The observational evidence of both the primordial and late time accelerating expansion of our Universe has stimulated the model building for inflation and dark energy as well as the exploration of theories of gravity beyond the general relativity (GR) (see \cite{Clifton:2011jh,Khoury:2013tda,Joyce:2014kja} for recent reviews).

Inflation and late time acceleration may be driven by some exotic matter content(s) of the Universe with unusual couplings to the gravity.
Along this line, significant progress has been made in rediscovering the Horndeski theory \cite{Horndeski:1974wa} --- the most general scalar-tensor theory involving derivatives up to the second order in the Lagrangian, while still leading to second order equations of motion for both scalar field and the metric --- as the ``generalized Galileon'' \cite{Deffayet:2011gz} (see \cite{Deffayet:2013lga,Charmousis:2014mia} for reviews).
Lagrangians for a single scalar field with nonlinear powers of second derivatives were systematically constructed in a Minkowski background  as the ``Galileon'' model \cite{Nicolis:2008in}, which was then generalized to a curved background using the  ``covariantization'' procedure \cite{Deffayet:2009wt,Deffayet:2009mn}.
The ``generalized Galileon'' \cite{Deffayet:2011gz} was constructed following the same procedure and
was  shown to be exactly equivalent to the Horndeski theory \cite{Kobayashi:2011nu}.
The ``second-order'' nature of Horndeski theory/generalized Galileon prevents it from extra ghostlike degrees of freedom and instabilities.

On the other hand, among various attempts to directly modify GR, massive gravity --- simply giving graviton a mass --- is a natural and simple choice.
The theory of free massive graviton considered by Fierz and Pauli (FP) \cite{Fierz:1939ix} does not recover GR in the massless
limit \cite{vanDam:1970vg,Zakharov:1970cc}. 
It was claimed that this discontinuity may be cured by nonlinearities \cite{Vainshtein:1972sx} in a possible full theory, although an arbitrary nonlinear generalization of the FP theory would inevitably excite the Boulware-Deser (BD)
ghost \cite{Boulware:1973my} --- an extra degree of freedom in addition to the usual five polarizations of a massive spin-2 particle.
Until recently, using the St\"{u}ckelberg method \cite{ArkaniHamed:2002sp}, consistent generalizations of the FP term were constructed by de Rham, Gabadadze, and Tolley (dRGT) \cite{deRham:2010ik,deRham:2010kj} (see \cite{deRham:2014zqa,Hinterbichler:2011tt,Rubakov:2008nh} for recent reviews). The dRGT theory possesses a Hamiltonian constraint (as well as a secondary constraint) necessary to remove the BD ghost \cite{Hassan:2011hr,Hassan:2011tf} (see also \cite{Deffayet:2012nr} for the constraint analysis in a covariant manner).
Nevertheless, the theoretical consistency of a massive gravity theory is still questioned \cite{Deser:2013rxa,Deser:2014hga,Deser:2014fta}.

Phenomenologically, it is interesting to couple the Galileon scalar field to massive gravity.
First, the original dRGT massive gravity with a flat fiducial metric only allows the existence of open FRW cosmology solutions under the homogeneous and isotropic gauge assumption \cite{deRham:2010tw,D'Amico:2011jj,Gumrukcuoglu:2011ew}, although spatially flat Friedmann-Robertson-Walker (FRW) solutions may exist when the gauge assumption or the fiducial metric itself is chosen to be nontrivial \cite{Chamseddine:2011bu,Gumrukcuoglu:2011zh,Hassan:2011vm,Kobayashi:2012fz,deRham:2012kf,Fasiello:2012rw,Langlois:2012hk,Langlois:2013cya,Gao:2014ula,Kugo:2014hja}.
Second, linear perturbations around FRW solutions of the original dRGT massive gravity suffer from instabilities \cite{DeFelice:2012mx,Khosravi:2013axa}.
One possible solution to these problems is to couple extra field(s) to dRGT massive gravity \cite{D'Amico:2012zv,Huang:2012pe,DeFelice:2013tsa,DeFelice:2013dua,Gumrukcuoglu:2013nza,D'Amico:2013kya,Gabadadze:2014kaa,Kahniashvili:2014wua} or to consider massive $f(R)$ theories \cite{Nojiri:2012zu,Kluson:2013yaa,Cai:2013lqa,Cai:2014upa,Wu:2014hva} (see also \cite{DeFelice:2013awa,deRham:2014gla} for other interesting attempts in order to have healthy cosmology in massive gravity).
Besides the physical metric, a fiducial metric is always present in Lorentz invariant mass terms. 
In particular, in the so-called bigravity theories \cite{Hassan:2011zd}, where the fiducial metric is also promoted to be dynamical, both metrics are treated in the same footing.
It is thus natural to explore how the extra fields can be coupled to both metrics (physical and fiducial) in massive gravity without introducing unwanted degree(s) of freedom such as the BD ghost \cite{Akrami:2013ffa,Tamanini:2013xia,Akrami:2014lja,Yamashita:2014fga,Noller:2014sta}.
If each matter field couples minimally to only one metric (physical or fiducial), it was proven that the BD ghost is absent at the classical level \cite{Hassan:2011zd}. 
It was further shown that this property still holds at the quantum level \cite{deRham:2014naa} (see e.g. \cite{Khosravi:2011zi,Akrami:2012vf,Akrami:2013ffa,Comelli:2014bqa,DeFelice:2014nja,Aoki:2014cla} for cosmological applications of such ``singly coupling'').
However, coupling the matter field(s) simultaneously to both metrics will not only  reintroduce the BD ghost at the classical level \cite{Yamashita:2014fga,deRham:2014naa,Soloviev:2014eea}, but also detune the dRGT potential by quantum loops \cite{deRham:2014naa}.
Nevertheless, a new type of doubly coupling to both metric was proposed in \cite{deRham:2014naa} (see also \cite{Heisenberg:2014rka}) in terms of a composite metric. It was claimed that the BD ghost is averted at all scales \cite{Hassan:2014gta}, while a further analysis shows that the BD ghost reappears but at a scale higher than the strong coupling scale \cite{deRham:2014fha}.
Comparing with the case where extra fields only couple to one metric, such doubly coupling with the composite metric seems to have promising cosmological applications \cite{Enander:2014xga,Schmidt-May:2014xla,Gumrukcuoglu:2014xba,Noller:2014ioa,Solomon:2014iwa,Mukohyama:2014rca}.

In this paper, we consider the Galileon field coupled to dRGT massive gravity through composite metrics.
In \cite{Gabadadze:2012tr,Andrews:2013ora,Andrews:2013uca,Goon:2014ywa}, the coupling between Galileon field(s) and massive gravity was introduced in a brane embedding manner. In this paper, we introduce the Galileon field and the composite metric in a straightforward way. 
We first derive the full set of equations of motion governing the evolution of a spatially flat Friedmann-Robertson-Walker (FRW) background.
Then we analyze the linear perturbations, including tensor, vector and scalar modes, around this background.
In particular, we would like to examine whether the nonminimal coupling between the Galileon field and massive gravity through the composite metric will reintroduce the BD ghost or not.

The rest of the present paper is organized as follows. 
In the next section we describe our model and notations. 
In Sec.\ref{sec:bgeom}, we derive the equations of motion for a FRW background.
In Sec.\ref{sec:pert} we derive the tensor, vector and scalar type perturbations around the FRW background, and make a brief analysis on their stability. 
Finally, we make a summary in Sec.\ref{sec:con}.
Throughout this paper, we set $c=M_{\mathrm{Pl}}=1$.

\section{Galileon coupled to massive gravity} \label{sec:model}

The dRGT theory \cite{deRham:2010ik,deRham:2010kj} describes nonderivative terms for metric perturbation $g_{\mu\nu}-f_{\mu\nu}$, where $g_{\mu\nu}$ is the physical metric, $f_{\mu\nu}$ is the covariantized fiducial metric given by
\begin{equation}
f_{\mu\nu}=\bar{f}_{ab}\frac{\partial\phi^{a}}{\partial x^{\mu}}\frac{\partial\phi^{b}}{\partial x^{\nu}}.
\end{equation}
Throughout this paper we consider spacetime to be four dimensional. Here $\{\phi^{a}\}\equiv\{\phi^0(x),\phi^i(x)\}$ are four St\"{u}ckelberg scalar fields, $\bar{f}_{ab}$ is the fixed metric in the field space. Generally the dimension of the St\"{u}ckelberg field space $\mathcal{N}$ may not necessarily be four dimensional. If $\mathcal{N}>4$, there will be $(\mathcal{N}-4)$ physical scalar degrees of freedom which cannot be gauged away. This is just the way the Galileon field was introduced and coupled to massive gravity in \cite{Gabadadze:2012tr,Andrews:2013ora,Andrews:2013uca,Goon:2014ywa}.
In this paper, we consider a flat fiducial metric, i.e., $\bar{f}_{ab} = \eta_{ab}$.
We consider the Galileon-type scalar field $\varphi$ coupled to dRGT massive gravity through a composite metric $g_{\mu\nu}(\alpha,\beta)$ introduced in \cite{deRham:2014naa}:
\begin{equation}
g_{\mu\nu}(\alpha,\beta)\equiv\alpha^{2}g_{\mu\nu}+2\alpha\beta\, g_{\mu\lambda}X_{\phantom{\lambda}\nu}^{\lambda}+\beta^{2}f_{\mu\nu}, \label{g_com}
\end{equation}
where $\alpha$, $\beta$ are numerical constants, and $X_{\phantom{\mu}\nu}^{\mu}$ is defined by
\begin{equation}
X_{\phantom{\mu}\lambda}^{\mu}X_{\phantom{\lambda}\nu}^{\lambda}\equiv g^{\mu\lambda}f_{\lambda\nu}.
\end{equation}

In this paper, we introduce the coupling between a Galileon scalar field and the dRGT massive gravity in a rather straightforward manner by considering the following  action:
\begin{equation}
S=S^{\mathrm{EH}}+S^{\mathrm{dRGT}}+S^{\mathrm{H}},\label{action}
\end{equation}
with
	\begin{equation}
		S^{\mathrm{EH}}  =  \frac{1}{2}\int\mathrm{d}^{4}x\sqrt{-g}R\left[g\right],\label{S_EH}
	\end{equation}
\begin{equation}
S^{\mathrm{dRGT}}  =  \frac{1}{2}\int\mathrm{d}^{4}x\sqrt{-g}\, m^{2}\left(e_{2}(\mathcal{\bm{K}})+\alpha_{3}e_{3}(\mathcal{\bm{K}})+\alpha_{4}e_{4}(\mathcal{\bm{K}})\right),\label{S_dRGT}
\end{equation}
where $R\left[g\right]$ is the Ricci scalar for the physical metric $g_{\mu\nu}$, $e_{2}\left(\mathcal{K}\right)$, $e_{3}\left(\mathcal{K}\right)$
and $e_{4}\left(\mathcal{K}\right)$ are dRGT potential terms with
\begin{equation}
\mathcal{K}_{\phantom{\mu}\nu}^{\mu}\equiv\delta_{\phantom{\mu}\nu}^{\mu}-X_{\phantom{\mu}\nu}^{\mu}.\label{K_def}
\end{equation}
Here and in the following, for a matrix $M_{\phantom{\mu}\nu}^{\mu}$, $e_{n}\left(\bm{M}\right)$
is defined by 
\begin{equation}
e_{n}\left(\bm{M}\right)\equiv M_{[\mu_{1}}^{\mu_{1}}M_{\mu_{2}}^{\mu_{2}}\cdots M_{\mu_{n}]}^{\mu_{n}},\label{sigma_n}
\end{equation}
where the antisymmetrization is \textit{unnormalized}. 
In four-dimensional spacetime, there are four independent Horndeski/Galileon terms \cite{Deffayet:2011gz}, which may generally have different composite metrics with different choices of $\alpha$ and $\beta$. For later convenience, we introduce four composite metrics, 
\begin{equation}
g_{\mu\nu}^{(i)}\equiv g_{\mu\nu}\big(\alpha^{(i)},\beta^{(i)}\big), \qquad i=0,1,2,3, \label{gi_def}
\end{equation}
and denote
\begin{eqnarray}
	X^{(i)} & \equiv &-\frac{1}{2}g^{(i)\mu\nu}\partial_{\mu}\varphi\partial_{\nu}\varphi,\nonumber\\
	\Pi_{\phantom{(i)\mu}\nu}^{(i)\mu} & \equiv & g^{(i)\mu\lambda}\nabla_{\lambda}^{(i)}\nabla_{\nu}^{(i)}\varphi,\nonumber\\
	i & = & 0,1,2,3,
\end{eqnarray}
where $g^{(i)\mu\nu}$ is the matrix inverse of $g^{(i)}_{\mu\nu}$, $\nabla^{(i)}_{\mu}$ is the covariant derivative compatible with $g^{(i)}_{\mu\nu}$.
The Horndeski/Galileon terms are thus given by 
\begin{equation}
S^{\mathrm{H}}= \sum_{i=0}^{3} \int\mathrm{d}t\mathrm{d}^{3}x\sqrt{-\det g^{(i)}_{\mu\nu}} \,\mathcal{L}^{(i)}, \label{S_H}
\end{equation}
with
\begin{equation}
\mathcal{L}^{(0)}  =  G^{(0)}\big(X^{(0)},\varphi\big),\label{LH0}
\end{equation}
\begin{equation}
\mathcal{L}^{(1)}  =  G^{(1)}\big(X^{(1)},\varphi\big)\, e_{1}\big(\bm{\Pi}^{(1)}\big),\label{LH1}
\end{equation}
\begin{equation}
\mathcal{L}^{(2)}  =  G^{(2)}\big(X^{(2)},\varphi\big)\, R\big[g^{(2)}\big]+\frac{\partial G^{(2)}}{\partial X^{(2)}}\, e_{2}\big(\bm{\Pi}^{(2)}\big),\label{LH2}
\end{equation}
\begin{equation}
\mathcal{L}^{(3)}  =  G^{(3)}\big(X^{(3)},\varphi\big)\, G^{\mu\nu}\big[g^{(3)}\big]\,\nabla_{\mu}^{(3)}\nabla_{\nu}^{(3)}\varphi-\frac{1}{6}\frac{\partial G^{(3)}}{\partial X^{(3)}}\, e_{3}\big(\bm{\Pi}^{(3)}\big),\label{LH3}
\end{equation}
where  $R[g^{(2)}]$ and $G^{\mu\nu}[g^{(3)}]$ denote the Ricci scalar and Einstein tensor for $g^{(2)}_{\mu\nu}$ and $g^{(3)}_{\mu\nu}$, respectively.

Note the Einstein-Hilbert term $R[g]$ can be absorbed into the Horndeski term $\mathcal{L}^{(2)}$ by choosing $G^{(2)}=1/2$ and $\alpha^{(2)}=1$ and $\beta^{(2)}=0$, while in this paper we keep the Einstein-Hilbert term for generality.

\subsection{Gauge fixing and variables}

We first fix three St\"{u}ckelberg fields $\{ \phi^{i}\}$ to be $\phi^{i}=a_{0}x^{i}$, where $a_{0}>0$ is some numerical constant, and write
\begin{equation}
\phi^{0}\left(t,\vec{x}\right)= \bar{\phi}\left(t\right)+\delta\phi\left(t,\vec{x}\right),\label{phi_0_exp}
\end{equation}
for short. 
For later convenience, we assume $\frac{\mathrm{d}{\bar{\phi}}}{\mathrm{d}t}>0$.
The Galileon scalar field $\varphi$ is perturbed as
\begin{equation}
\varphi=\bar{\varphi}\left(t\right)+\delta\varphi\left(t,\vec{x}\right).\label{varphi_exp}
\end{equation}
We are still left with one gauge degree of freedom.
We may further fix $\delta\phi\left(t,\vec{x}\right)=0$  or $\delta\varphi\left(t,\vec{x}\right)=0$.
In this paper, we choose $\delta\phi\left(t,\vec{x}\right)=0$ which we may refer to as  the ``almost unitary gauge.''

For the physical metric, it is convenient to work with ADM variables defined by
\begin{equation}
\mathrm{d}s_{g}^{2} = -N^{2}dt^{2}+h_{ij}\left(dx^{i}+N^{i}dt\right)\left(dx^{j}+N^{j}dt\right),
\end{equation}
where $\left\{ N,N_{i},h_{ij}\right\} $ are parametrized by
\begin{equation}
N = \bar{N}e^{A},
\end{equation}
\begin{equation}
N_{i}  =  \bar{N}aB_{i},
\end{equation}
\begin{equation}
h_{ij}  =  a^{2}(e^{\bm{H}})_{ij}\equiv a^{2}\bigg(\delta_{ij}+H_{ij}+\frac{1}{2}H_{ik}H_{kj}+\cdots\bigg),
\end{equation}
with $\bar{N}=\bar{N}\left(t\right)$, $a=a\left(t\right)$, and $B_i$ and $H_{ij}$ can be further decomposed as ($\partial^{2}\equiv\delta^{ij}\partial_{i}\partial_{j}$)
\begin{equation}
B_{i} \equiv \partial_{i}B+S_{i},
\end{equation}
\begin{equation}
H_{ij}  \equiv  2\zeta\delta_{ij}+\left(\partial_{i}\partial_{j}-\frac{1}{3}\delta_{ij}\partial^{2}\right)E+\partial_{(i}F_{j)}+\gamma_{ij}.
\end{equation}
We require
\begin{equation}
\partial_{i}S_{i}=\partial_{i}F_{i}=0,\qquad\partial_{i}\gamma_{ij}=0,\qquad\gamma_{ii}=0.
\end{equation}
By definition, the covariantized fiducial metric $f_{\mu\nu}$ is unperturbed in the almost unitary gauge, i.e.,
\begin{equation}
f_{\mu\nu}\rightarrow\left(\begin{array}{cc}
-\left(\partial_{t}\bar{\phi}\right)^{2} & 0\\
0 & a_{0}^{2}\delta_{ij}
\end{array}\right).
\end{equation}
At the background level, we have
\begin{equation}
g_{(0)\mu\nu}\rightarrow\left(\begin{array}{cc}
-\bar{N}^{2} & 0\\
0 & a^{2}\delta_{ij}
\end{array}\right),
\end{equation}
and thus for each composite metric in (\ref{gi_def}),
\begin{equation}
g_{(0)\mu\nu}^{(i)}\rightarrow\left(\begin{array}{cc}
-\left(N^{(i)}\right)^{2} & 0\\
0 & \left(a^{(i)}\right)^{2}\delta_{ij}
\end{array}\right),\qquad i=0,1,2,3,
\end{equation}
with
\begin{eqnarray}
N^{(i)} & \equiv & \alpha^{(i)}\bar{N}+\beta^{(i)}\frac{\mathrm{d}\bar{\phi}}{\mathrm{d}t},\nonumber\\
a^{(i)} & \equiv & \alpha^{(i)}a+\beta^{(i)}a_{0},\nonumber\\
i & = & 0,1,2,3. \label{Ni_ai}
\end{eqnarray}

At this point, apparently we have
\begin{eqnarray*}
	\text{five scalar modes:} &  & A,\quad B,\quad\zeta,\quad E,\quad\delta\varphi,\\
	\text{four vector modes:} &  & S_{i},\quad F_{i},\\
	\text{two tensor modes:} &  & \gamma_{ij}.
\end{eqnarray*}
As we shall see below, at the level of linear perturbations, only two (out of four) independent vector modes are propagating, while the nonminimal coupling between the Galileon field and the massive gravity through composite metrics will excite three (out of five) independent scalar modes, one of which may correspond to the BD ghost.

\section{Background equations of motion} \label{sec:bgeom}

When being perturbed around a FRW background, the background equations of motion are determined by requiring the vanishing of the first order Lagrangian, which takes the form
\begin{equation}
\mathcal{L}_{1}=\int\mathrm{d}t\mathrm{d}^{3}x\left[\bar{N}a^{3}\left(\mathcal{E}_{A}A+3\mathcal{E}_{\zeta}\zeta\right)+\mathcal{E}_{\delta\varphi}\delta\varphi\right] = 0,
\end{equation}
with
\begin{equation}
\mathcal{E}_{A} \equiv 3H^{2}-\rho,\label{eom_A}
\end{equation}
\begin{equation}
\mathcal{E}_{\zeta} \equiv 3H^{2}+2\dot{H}+P,\label{eom_zeta}
\end{equation}
\begin{equation}
\mathcal{E}_{\delta\varphi} \equiv \sum_{i=0}^{3}N^{(i)}\big(a^{(i)}\big)^{3}\bigg[\bar{\mathcal{L}}_{,\varphi}^{(i)}-\frac{1}{N^{(i)}(a^{(i)})^{3}}\frac{\mathrm{d}}{\mathrm{d}t}\Big(\big(a^{(i)}\big)^{3}\mathcal{J}^{(i)}\Big)\bigg],\label{eom_varphi}
\end{equation}
where $N^{(i)}$ and $a^{(i)}$ are defined in (\ref{Ni_ai}), and throughout this paper, for any quantity $q$ we denote
\begin{equation}
\dot{q}\equiv\frac{1}{\bar{N}}\frac{\partial q}{\partial t},\qquad\ddot{q}\equiv\frac{1}{\bar{N}}\frac{\partial}{\partial t}\left(\frac{1}{\bar{N}}\frac{\partial q}{\partial t}\right),
\end{equation}
for short and the Hubble parameter is defined to be
\begin{equation}
H \equiv \frac{1}{\bar{N}a}\frac{\mathrm{d}a}{\mathrm{d}t} \equiv\frac{\dot{a}}{a}.
\end{equation}
In (\ref{eom_A}) and (\ref{eom_zeta}) $\rho$ and $P$ are effective energy density and pressure respectively, of which the explicit expressions will be given below.
Recall that in the usual case of GR with a scalar field, among three
equations of motion
\begin{equation}
\mathcal{E}_{A}=0,\qquad\mathcal{E}_{\zeta}=0,\qquad\mathcal{E}_{\delta\varphi}=0,
\end{equation}
only two of them are independent. However in the case of massive gravity, all three equations are independent.

For later convenience, we also define
\begin{eqnarray}
\dot{q}^{(i)} & \equiv & \frac{1}{N^{(i)}}\frac{\partial q}{\partial t},\nonumber\\
\ddot{q}^{(i)} & \equiv & \frac{1}{N^{(i)}}\frac{\partial}{\partial t}\left(\frac{1}{N^{(i)}}\frac{\partial q}{\partial t}\right),\nonumber\\ 
i & =& 0,1,2,3, \label{dtddti}
\end{eqnarray}
etc., and
\begin{equation}
H^{(i)}\equiv\frac{1}{N^{(i)}a^{(i)}}\frac{\mathrm{d}a^{(i)}}{\mathrm{d}t}.
\end{equation}
Using these notations, at the background level, we may write
\begin{equation}
X^{(i)}\equiv-\frac{1}{2}g_{(0)}^{(i)\mu\nu}\left(\partial_{t}\bar{\varphi}\right)^{2}=\frac{\left(\partial_{t}\bar{\varphi}\right)^{2}}{2\left(N^{(i)}\right)^{2}}\equiv\frac{1}{2}\big(\dot{\bar{\varphi}}^{(i)}\big)^{2}.\label{Xi_bg}
\end{equation}

After some manipulations, the effective energy density $\rho$ in (\ref{eom_A}) is given by
\begin{equation}
\rho=\rho_{m}+\sum_{i=0}^{3}\rho^{(i)},
\end{equation}
with
\begin{align}
\rho_{m}  = &\; -3m^{2}\bigg[2\left(1+2\alpha_{3}+2\alpha_{4}\right)-3\frac{a_{0}}{a}\left(1+3\alpha_{3}+4\alpha_{4}\right)\nonumber \\
&   +\Big(\frac{a_{0}}{a}\Big)^{2}(1+6\alpha_{3}+12\alpha_{4})-\Big(\frac{a_{0}}{a}\Big)^{3}(\alpha_{3}+4\alpha_{4})\bigg],\qquad \label{rho_m}
\end{align}
and
\begin{equation}
\rho^{(0)}  =  -\alpha^{(0)}\left(\frac{a^{(0)}}{a}\right)^{3}\left(G^{(0)}-2X^{(0)}G_{,X}^{(0)}\right),\label{rho_0}
\end{equation}
\begin{equation}
\rho^{(1)}  =  -2\alpha^{(1)}\left(\frac{a^{(1)}}{a}\right)^{3}\Big(3H^{(1)}\dot{\bar{\varphi}}^{(1)}X^{(1)}G_{,X}^{(1)}-X^{(1)}G_{,\varphi}^{(1)}\Big),\label{rho_1}
\end{equation}
\begin{align}
\rho^{(2)} = & -6\alpha^{(2)}\left(\frac{a^{(2)}}{a}\right)^{3}\Big[\left(H^{(2)}\right)^{2}\Big(G^{(2)}-4X^{(2)}G_{,X}^{(2)} \nonumber\\
& -4(X^{(2)})^{2}G_{,XX}^{(2)}\Big)+H^{(2)}\dot{\bar{\varphi}}^{(2)}\big(2X^{(2)}G_{,X\varphi}^{(2)}+G_{,\varphi}^{(2)}\big)\Big],\label{rho_2}
\end{align}
\begin{align}
\rho^{(3)}  = & -\alpha^{(3)}\left(\frac{a^{(3)}}{a}\right)^{3}\Big[-(H^{(3)})^{3}\dot{\bar{\varphi}}^{(3)}X^{(3)}\big(10G_{,X}^{(3)}  \nonumber\\
&+4X^{(3)}G_{,XX}^{(3)}\big)  +(H^{(3)})^{2}X^{(3)}\big(12X^{(3)}G_{,X\varphi}^{(3)}+18G_{,\varphi}^{(3)}\big)\Big],\label{rho_3}
\end{align}
where we used the shorthands $G_{,X}^{(0)}\equiv\frac{\partial G^{(0)}}{\partial X^{(0)}}$ and $G_{,X\varphi}^{(2)}\equiv\frac{\partial^{2}G^{(2)}}{\partial X^{(2)}\partial\varphi}$ etc., and $\dot{\bar{\varphi}}^{(1)}$ etc. are defined as (\ref{dtddti}).
Comparing with the effective energy density of usual Galileon field (e.g. \cite{Gao:2011qe,Gao:2011mz,DeFelice:2011uc}), $\rho^{(i)}$'s contain an additional factor $\big(a^{(i)}/a)^3$. In the case $\beta^{(i)}=0$, i.e. the Galileon field couples to massive gravity only through the physical metric, $a^{(i)}/a = \alpha^{(i)}$ and thus the corresponding energy density $\rho^{(i)}$ is simply proportional to the energy density of the usual Galileon field.
Conversely, if $\alpha^{(i)}=0$, i.e. the Galileon field couples to massive gravity only through the fiducial metric $f_{\mu\nu}$, we have $\big(a^{(i)}/a)^3 \sim 1/a^3$, which implies the contributions to the effective energy density from the Galileon field are redshifted as $1/a^3$.

The effective pressure $P$ in (\ref{eom_zeta}) is given by
\begin{equation}
P=P_{m}+\sum_{i=0}^{3}P^{(i)}, \label{P_def}
\end{equation}
with
\begin{align}
P_{m} = &\; 3m^{2}\left[2\left(1+2\alpha_{3}+2\alpha_{4}\right)-\dot{\bar{\phi}}\left(1+3\alpha_{3}+4\alpha_{4}\right)\right]\nonumber \\
&   -2m^{2}\frac{a_{0}}{a}\left[3\left(1+3\alpha_{3}+4\alpha_{4}\right)-\dot{\bar{\phi}}\left(1+6\alpha_{3}+12\alpha_{4}\right)\right]\nonumber \\
&   +m^{2}\left(\frac{a_{0}}{a}\right)^{2}\left[1+6\alpha_{3}+12\alpha_{4}-3\dot{\bar{\phi}}\left(\alpha_{3}+4\alpha_{4}\right)\right],\label{P_m}
\end{align}
\begin{widetext}
\begin{eqnarray}
P^{(0)} & = & \alpha^{(0)}\frac{N^{(0)}}{\bar{N}}\left(\frac{a^{(0)}}{a}\right)^{2}G^{(0)},\label{P_0}\\
P^{(1)} & = & 2\alpha^{(1)}\frac{N^{(1)}}{\bar{N}}\left(\frac{a^{(1)}}{a}\right)^{2}X^{(1)}\left(G_{,\varphi}^{(1)}+G_{,X}^{(1)}\ddot{\bar{\varphi}}^{(1)}\right),\label{P_1}
\end{eqnarray}
\begin{eqnarray}
P^{(2)} & = & \alpha^{(2)}\frac{N^{(2)}}{\bar{N}}\left(\frac{a^{(2)}}{a}\right)^{2}\Big\{2G_{,\varphi}^{(2)}\left(\ddot{\bar{\varphi}}^{(2)}+2H^{(2)}\dot{\bar{\varphi}}^{(2)}\right)-8H^{(2)}\dot{\bar{\varphi}}^{(2)}\ddot{\bar{\varphi}}^{(2)}X^{(2)}G_{,XX}^{(2)}\nonumber \\
&  & +4X^{(2)}G_{,\varphi\varphi}^{(2)}+4X^{(2)}G_{,X\varphi}^{(2)}\left(\ddot{\bar{\varphi}}^{(2)}-2H^{(2)}\dot{\bar{\varphi}}^{(2)}\right)+2G^{(2)}\left(3\big(H^{(2)}\big)^{2}+2\dot{H}^{(2)}\right)\nonumber \\
&  & -4G_{,\tilde{X}}^{(2)}\left[\left(3\big(H^{(2)}\big)^{2}+2\dot{H}^{(2)}\right)\tilde{X}+H^{(2)}\dot{\bar{\varphi}}^{(2)}\ddot{\bar{\varphi}}^{(2)}\right]\Big\},\label{P2}
\end{eqnarray}
and
\begin{eqnarray}
P^{(3)} & = & \alpha^{(3)}\frac{N^{(3)}}{\bar{N}}\left(\frac{a^{(3)}}{a}\right)^{2}\Big\{-4\big(H^{(3)}\big)^{2}\big(X^{(3)}\big)^{2}G_{,XX}^{(3)}\ddot{\bar{\varphi}}^{(2)}+4H^{(3)}\dot{\bar{\varphi}}^{(3)}X^{(3)}G_{,\varphi\varphi}^{(3)}\nonumber \\
&  & -2H^{(3)}X^{(3)}G_{,X}^{(3)}\left[2\left(\dot{H}^{(3)}+\big(H^{(3)}\big)^{2}\right)\dot{\bar{\varphi}}^{(3)}+3H^{(3)}\ddot{\bar{\varphi}}^{(3)}\right]\nonumber \\
&  & -4H^{(3)}X^{(3)}G_{,X\varphi}^{(3)}\left(H^{(3)}X^{(3)}-\dot{\bar{\varphi}}^{(3)}\ddot{\bar{\varphi}}^{(3)}\right)\nonumber \\
&  & +2G_{,\varphi}^{(3)}\left[\left(3\big(H^{(3)}\big)^{2}+2\dot{H}^{(3)}\right)X^{(3)}+2H^{(3)}\dot{\bar{\varphi}}^{(3)}\ddot{\bar{\varphi}}^{(3)}\right]\Big\},\label{P3}
\end{eqnarray}
\end{widetext}
where $\dot{H}^{(2)}\equiv\frac{1}{N^{(2)}}\frac{\mathrm{d}H^{(2)}}{\mathrm{d}t}$ and $\dot{H}^{(3)}\equiv\frac{1}{N^{(3)}}\frac{\mathrm{d}H^{(3)}}{\mathrm{d}t}$.
Similar to the analysis for the energy density, due to the factor $\big(a^{(i)}/a)^2$, the contributions to the effective pressure from the Galileon field will redshift as $1/a^2$  if the scalar field only couples to the fiducial metric.

For the background equations of motion for the scalar field $\varphi$ (\ref{eom_varphi}), we have
\begin{equation}
\bar{\mathcal{L}}_{,\varphi}^{(0)}  =  G_{,\varphi}^{(0)},\label{Lvp0}
\end{equation}
\begin{equation}
\bar{\mathcal{L}}_{,\varphi}^{(1)}  =  -G_{,\varphi}^{(1)}\frac{1}{N^{(1)}\left(a^{(1)}\right)^{3}}\frac{\mathrm{d}}{\mathrm{d}t}\left(\big(a^{(1)}\big)^{3}\dot{\bar{\varphi}}^{(1)}\right),\label{Lvp1}
\end{equation}
\begin{align}
\bar{\mathcal{L}}_{,\varphi}^{(2)}  = &\; 6\left(\dot{H}^{(2)}+2\big(H^{(2)}\big)^{2}\right)G_{,\varphi}^{(2)} \nonumber\\
&+6H^{(2)}\left(\dot{\bar{\varphi}}^{(2)}\ddot{\bar{\varphi}}^{(2)}+2X^{(2)}H^{(2)}\right)G_{,X\varphi}^{(2)},\label{Lvp2}
\end{align}
\begin{align}
\bar{\mathcal{L}}_{,\varphi}^{(3)}  = &\; 2\big(H^{(3)}\big)^{2}\left(3\ddot{\bar{\varphi}}^{(3)}+H^{(3)}\dot{\bar{\varphi}}^{(3)}\right)X^{(3)}G_{,X\varphi}^{(3)}\nonumber \\
&   +3H^{(3)}\Big(\big(3\big(H^{(3)}\big)^{2}+2\dot{H}^{(3)}\big)\dot{\bar{\varphi}}^{(3)}+H^{(3)}\ddot{\bar{\varphi}}^{(3)}\Big)G_{,\varphi}^{(3)},\label{Lvp3}
\end{align}
and
\begin{equation}
\mathcal{J}^{(0)}  =  \dot{\bar{\varphi}}^{(0)}G_{,X}^{(0)},\label{J0}
\end{equation}
\begin{equation}
\mathcal{J}^{(1)}  =  \dot{\bar{\varphi}}^{(1)}G_{,\varphi}^{(1)}-6H^{(1)}X^{(1)}G_{,X}^{(1)},\label{J1}
\end{equation}
\begin{align}
\mathcal{J}^{(2)}  = &\; 6\big(H^{(2)}\big)^{2}\dot{\bar{\varphi}}^{(2)}G_{,X}^{(2)}+12\big(H^{(2)}\big)^{2}\dot{\bar{\varphi}}^{(2)}\tilde{X}G_{,\tilde{X}\tilde{X}}^{(2)} \nonumber\\
& -12H^{(2)}X^{(2)}G_{,X\varphi}^{(2)},\label{J2}
\end{align}
\begin{align}
\mathcal{J}^{(3)}  = &\; 6\big(H^{(3)}\big)^{3}X^{(3)}G_{,X}^{(3)}+4\big(H^{(3)}\big)^{3}\big(X^{(3)}\big)^{2}G_{,XX}^{(3)}\nonumber \\
&   -3\big(H^{(3)}\big)^{2}\dot{\bar{\varphi}}^{(3)}G_{,\varphi}^{(3)}-6\big(H^{(3)}\big)^{2}\dot{\bar{\varphi}}^{(3)}X^{(3)}G_{,X\varphi}^{(3)}.\label{J3}
\end{align}

\section{Linear perturbations} \label{sec:pert}

One motivation to introduce the composite metric (\ref{g_com}) in \cite{deRham:2014naa} is to have simultaneous coupling of the extra fields to both physical and fiducial metrics without introducing the BD ghost. 
In this paper we would like to examine if such a composite metric can be applied to the Galileon scalar field with more general nonminimal/kinetic couplings to gravity. In particular, we are interested in whether the BD ghost reappears or not, at least at the level of linear perturbations around a FRW background.

\subsection{Tensor modes}

The quadratic Lagrangian for the tensor modes $\gamma_{ij}$ takes the form (in Fourier space)
\begin{eqnarray}
S_{2}\left[\gamma_{ij}\right] & = & \frac{1}{8}\int\mathrm{d}t\frac{\mathrm{d}^{3}k}{\left(2\pi\right)^{3}}\,\bar{N}a^{3}\Big[\left(1+g_{\gamma\gamma}\right)\dot{\gamma}_{ij}^{2} \nonumber\\
&& -\left(1+w_{\gamma\gamma}\right)\frac{k^{2}}{a^{2}}\gamma_{ij}^{2}-m_{\gamma\gamma}\gamma_{ij}^{2}\Big], \label{S2_ten}
\end{eqnarray}
where terms such as $\gamma_{ij}^2$ are shorthands for $\gamma_{ij}(t,\bm{k}) \gamma_{ij}(t,-\bm{k})$ etc. Please keep in mind that $\dot{\gamma}_{ij}\equiv \frac{1}{\bar{N}}\partial_t \gamma_{ij}$. 
In (\ref{S2_ten}), various coefficients are given by
\begin{equation}
g_{\gamma\gamma}=g_{\gamma\gamma}^{(2)}+g_{\gamma\gamma}^{(3)},
\end{equation}
with
\begin{eqnarray}
g_{\gamma\gamma}^{(2)} & = & 2\big(\alpha^{(2)}\big)^{2}\frac{b^{(2)}}{b}\left(G^{(2)}-2X^{(2)}G_{,X}^{(2)}\right),\label{g_gg2}\\
g_{\gamma\gamma}^{(3)} & = & 2\big(\alpha^{(3)}\big)^{2}\frac{b^{(3)}}{b}X^{(3)}\left(G_{,\varphi}^{(3)}-H^{(3)}\dot{\bar{\varphi}}^{(3)}G_{,X}^{(3)}\right),\label{g_gg3}
\end{eqnarray}
and
\begin{align}
w_{\gamma\gamma}  = &\; 2\big(\alpha^{(2)}\big)^{2}\frac{b}{b^{(2)}}G^{(2)} \nonumber\\
&-2\big(\alpha^{(3)}\big)^{2}\frac{b}{b^{(3)}}X^{(3)}\left(G_{,\varphi}^{(3)}+G_{,X}^{(3)}\ddot{\bar{\varphi}}^{(3)}\right),\label{w_gg}
\end{align}
\begin{equation}
m_{\gamma\gamma}  =  -\frac{1}{2}a_{0}\frac{\partial P}{\partial a_{0}},\label{m_gg}
\end{equation}
with $P$ the pressure given in (\ref{P_def}) and
\begin{equation}
b\equiv\frac{a}{\bar{N}a_{0}}\frac{\mathrm{d}\bar{\phi}}{\mathrm{d}t},\qquad b^{(i)}\equiv\frac{a^{(i)}}{N^{(i)}a_{0}}\frac{\mathrm{d}\bar{\phi}}{\mathrm{d}t},\qquad i=0,1,2,3.
\end{equation}
Please note $\frac{\partial P^{(2)}}{\partial a_{0}}$ and $\frac{\partial P^{(3)}}{\partial a_{0}}$ also receive contributions from $H^{(2)}$ and $H^{(3)}$ since $\frac{\partial H^{(i)}}{\partial a_{0}}=-\frac{\beta^{(i)}}{a^{(i)}}H^{(i)}$.

Note $g_{\gamma\gamma}$ and $w_{\gamma\gamma}$ are nonvanishing only in the presence of $\mathcal{L}^{(2)}$ and (or) $\mathcal{L}^{(3)}$. In this case, the tensor modes propagate with a modified speed of sound
\begin{equation}
c_{\mathrm{T}}^2 = \frac{1+w_{\gamma\gamma}}{1+g_{\gamma\gamma}} .
\end{equation}
In order to ensure that the tensor modes have the right sign for the kinetic term and do not have gradient instabilities, we must have both $1+g_{\gamma\gamma}>0$ and $1+w_{\gamma\gamma}>0$.
From (\ref{m_gg}), it is interesting to note that 
\begin{equation}
\frac{\partial P^{(i)}}{\partial a_{0}}\neq0,\qquad\text{if and only if}\quad\beta^{(i)}\neq0,\quad i=0,1,2,3,
\end{equation}
which implies that even without the dRGT potential terms, the linear gravitational waves will become massive due to the presence of the composite metric $g^{(i)}_{\mu\nu}$ in the Horndeski/Galileon terms.

\subsection{Vector modes}

The quadratic Lagrangian for the vector modes is given by
\begin{eqnarray}
S_{2}\left[B_{i},F_{i}\right] & = & \frac{1}{4}\int\mathrm{d}t\frac{\mathrm{d}^{3}k}{\left(2\pi\right)^{3}}\,\bar{N}a^{3}\Big[\left(1+w_{BB}\right)\frac{k^{2}}{a^{2}}B_{i}B_{i} \nonumber\\
& & +m_{BB}\, B_{i}B_{i} -\left(1+f_{BF}\right)\frac{k^{2}}{a}B_{i}\dot{F}_{i} \nonumber\\
&& -w_{BF}\frac{k^{2}}{a}B_{i}F_{i} +\left(1+g_{\gamma\gamma}\right)\frac{k^{2}}{4}\dot{F}_{i}\dot{F}_{i} \nonumber\\
&& -m_{\gamma\gamma}\frac{k^{2}}{4}F_{i}F_{i}\Big],\label{S2_vec}
\end{eqnarray}
with
\begin{equation}
w_{BB} =  \sum_{i=2}^{3}\left(\frac{1+b^{(i)}}{1+b}\frac{b}{b^{(i)}}\right)^{2}g_{\gamma\gamma}^{(i)},\label{w_BB}
\end{equation}
\begin{align}
m_{BB}  = &\; \frac{2}{3}\frac{1}{\left(1+b\right)}a_{0}\frac{\partial\rho_{m}}{\partial a_{0}} +\frac{1}{\left(1+b\right)^{2}}2\sum_{i=0}^{3}\beta^{(i)}\frac{a_{0}}{a^{(i)}} \nonumber\\
&\times \left[P^{(i)}+\left(1+b+\frac{b}{b^{(i)}}\right)\rho^{(i)}\right],\label{m_BB}
\end{align}
\begin{equation}
f_{BF}  =  \sum_{i=2}^{3}\frac{1+b^{(i)}}{1+b}\frac{b}{b^{(i)}}g_{\gamma\gamma}^{(i)},\label{f_BB}
\end{equation}
\begin{equation}
w_{BF}  =  H\sum_{i=2}^{3}\beta^{(i)}\frac{a_{0}}{a^{(i)}}\frac{1+b^{(i)}}{1+b}\frac{b}{b^{(i)}}g_{\gamma\gamma}^{(i)},\label{w_BF}
\end{equation}
where $g_{\gamma\gamma}^{(i)}$ are given in (\ref{g_gg2}) and (\ref{g_gg3}).
At this point, note $m_{BB}\neq 0$ only if at least one $\beta^{(i)}\neq 0$.

In (\ref{S2_vec}) $B_i$ plays as an auxiliary variable since (\ref{S2_vec}) contains no time derivative of it. Thus $B_i$ can be solved in terms of $F_i$ and $\dot{F}_i$ by varying (\ref{S2_vec}) with respect to $B_{i}$, which yields
\begin{equation}
B_{i}=\frac{1}{2}\frac{1}{\left(1+w_{BB}\right)\frac{k^{2}}{a^{2}}+m_{BB}}\frac{k^{2}}{a}\Big(\left(1+f_{BF}\right)\dot{F}_{i}+w_{BF}F_{i}\Big).\label{Bi_sol}
\end{equation}
Plugging (\ref{Bi_sol}) into (\ref{S2_vec}), we get the quadratic action for $F_i$ alone:
\begin{equation}
S_{2}\left[F_{i}\right]=\frac{1}{4}\int\mathrm{d}t\frac{\mathrm{d}^{3}k}{\left(2\pi\right)^{3}}\,\bar{N}a^{3}\frac{k^{2}}{4}\left(\mathcal{G}_{F}\dot{F}_{i}^{2}-\mathcal{W}_{F}F_{i}^{2}\right),
\end{equation}
with
\begin{equation}
\mathcal{G}_{F}  =  1+g_{\gamma\gamma}-\frac{\left(1+f_{BF}\right)^{2}}{\left(1+w_{BB}\right)\frac{k^{2}}{a^{2}}+m_{BB}}\frac{k^{2}}{a^{2}},\label{GF}
\end{equation}
\begin{align}
\mathcal{W}_{F}  = &\; m_{\gamma\gamma}+\frac{k^{2}}{a^{2}}\bigg[\frac{w_{BF}^{2}}{\left(1+w_{BB}\right)\frac{k^{2}}{a^{2}}+m_{BB}} \nonumber\\
& +\frac{1}{\bar{N}a}\frac{\mathrm{d}}{\mathrm{d}t}\bigg(\frac{a\left(1+f_{BF}\right)w_{BF}}{\left(1+w_{BB}\right)\frac{k^{2}}{a^{2}}+m_{BB}}\bigg)\bigg].\label{WF}
\end{align}
In order to avoid instabilities for $F_i$ we must require $\mathcal{G}_{F}>0$ and $\mathcal{W}_F>0$.

\subsection{Scalar modes}

As we have described before, the scalar-type perturbations around a FRW background are characterized by five variables $A$, $B$, $E$, $\zeta$ and $\delta\varphi$. 
Straightforward calculation yields the quadratic Lagrangian:
\begin{equation}
S_{2}\left[A,B,E,\zeta,\delta\varphi\right]=\int\mathrm{d}t\frac{\mathrm{d}^{3}k}{\left(2\pi\right)^{3}}\bar{N}a^{3}\mathcal{L}_{2}^{\mathrm{(s)}}, \label{S2_S_ori}
\end{equation}
with
\begin{widetext}
\begin{eqnarray}
\mathcal{L}_{2}^{\mathrm{(s)}} & = & -\left(1+m_{AA}\right)\frac{3}{2}H^{2}A^{2}+\frac{1}{4}m_{BB}k^{2}B^{2}+\left(1+g_{\gamma\gamma}\right)\frac{k^{4}}{12}\dot{E}^{2} +\frac{1}{12}\left(\frac{1}{3}\left(1+w_{\gamma\gamma}\right)\frac{k^{2}}{a^{2}}-m_{\gamma\gamma}\right)k^{4}E^{2}-\left(1+g_{\gamma\gamma}\right)3\dot{\zeta}^{2}\nonumber \\
&  & +\left[\left(1+w_{\gamma\gamma}\right)\frac{k^{2}}{a^{2}}-\frac{3}{2}a_{0}\frac{\partial P}{\partial a_{0}}\right]\zeta^{2}+g_{\varphi\varphi}\delta\dot{\varphi}^{2}+\left(w_{\varphi\varphi}\frac{k^{2}}{a^{2}}+m_{\varphi\varphi}\right)\delta\varphi^{2} +\left(1+w_{AB}\right)2Hk^{2}A\frac{B}{a}+\left(1+w_{AE}\right)\frac{1}{3}\frac{k^{4}}{a^{2}}AE\nonumber\\
&  & +\left(1+f_{A\zeta}\right)6HA\dot{\zeta} +\left(\left(1+w_{A\zeta}\right)2\frac{k^{2}}{a^{2}}+a_{0}\frac{\partial\rho}{\partial a_{0}}\right)A\zeta+f_{A\varphi}A\delta\dot{\varphi}+\left(w_{A\varphi}\frac{k^{2}}{a^{2}}-\frac{\partial\rho}{\partial\varphi}\right)A\delta\varphi -\left(1+f_{BE}\right)\frac{k^{4}}{3}\frac{B}{a}\dot{E} \nonumber\\
&  & +w_{BE}k^{4}\frac{B}{a}E-\left(1+f_{BE}\right)2k^{2}\frac{B}{a}\dot{\zeta}+6w_{BE}k^{2}\frac{B}{a}\zeta +f_{B\varphi}k^{2}\frac{B}{a}\delta\dot{\varphi}  +w_{B\varphi}k^{2}\frac{B}{a}\delta\varphi+\left(1+w_{\gamma\gamma}\right)\frac{1}{3}\frac{k^{4}}{a^{2}}E\zeta+\frac{1}{6}w_{\zeta\varphi}\frac{k^{4}}{a^{2}}E\delta\varphi\nonumber \\
&  & +g_{\zeta\varphi}\dot{\zeta}\delta\dot{\varphi}+f_{\zeta\varphi}\zeta\delta\dot{\varphi}+\left(w_{\zeta\varphi}\frac{k^{2}}{a^{2}}+3\frac{\partial P}{\partial\varphi}\right)\zeta\delta\varphi,\label{L2_s_ori}
\end{eqnarray}
\end{widetext}
where various coefficients are listed in Appendix \ref{sec:coeff}.

Please note (\ref{L2_s_ori}) contains no time derivative of $A$ nor $B$, and thus we may solve $A$ and $B$ in terms of other three variables $\zeta$, $E$ and $\delta\varphi$. A full treatment is out of the scope of this paper. Here we concentrate on the question: Will the composite metric appearing in the Horndeski/Galileon Lagrangians reintroduce the BD ghost or not?

To this end, we first solve the constraint equations for $A$ and $B$ with the following solutions:
\begin{widetext}
\begin{eqnarray}
A & = & \frac{1}{H\Xi}\Big\{\frac{4}{3}k^{4}\left(1+f_{BE}\right)\left(1+w_{AB}\right)\dot{E}+\left[a^{2}f_{A\ensuremath{\varphi}}m_{BB}-4Hk^{2}f_{B\ensuremath{\varphi}}\left(1+w_{AB}\right)\right]\frac{\delta\dot{\varphi}}{H}\nonumber \\
&  & +\left[6a^{2}\left(1+f_{A\ensuremath{\zeta}}\right)m_{BB}+8k^{2}\left(1+f_{BE}\right)\left(1+w_{AB}\right)\right]\dot{\zeta}\Big\}+\cdots,\label{A_sol}\\
\frac{B}{a} & = & \frac{1}{\Xi}\Big\{2k^{2}\left(1+f_{BE}\right)\left(1+m_{AA}\right)\dot{E}-2\left[3Hf_{B\ensuremath{\varphi}}\left(1+m_{AA}\right)+2f_{A\ensuremath{\varphi}}\left(1+w_{AB}\right)\right]\frac{\delta\dot{\varphi}}{H}\nonumber \\
&  & +12\left[-2f_{A\ensuremath{\zeta}}\left(1+w_{AB}\right)+f_{BE}m_{AA}+f_{BE}+m_{AA}-2w_{AB}-1\right]\dot{\zeta}\Big\}+\cdots,\label{B_sol}
\end{eqnarray}
\end{widetext}
with
\begin{equation}
\Xi\equiv3a^{2}\left(1+m_{AA}\right)m_{BB}+8k^{2}\left(1+w_{AB}\right)^{2}, \label{Xi_def}
\end{equation}
where ``$\cdots$'' denote terms which do not contribute to the kinetic terms of $E$, $\delta\varphi$ and $\zeta$.
By plugging (\ref{A_sol}) and (\ref{B_sol}) into (\ref{L2_s_ori}) and concentrating on the kinetic terms, we get
\begin{equation}
\mathcal{L}_{2}^{\mathrm{(s)}}[E,\zeta,\delta\varphi]=\frac{1}{2}\dot{q}^{\mathrm{T}}\,\bm{G}\,\dot{q}+\cdots, \quad \text{with } q^{\mathrm{T}}\equiv\left(\begin{array}{ccc}
E & \zeta & \delta\varphi\end{array}\right), \label{L2s_kin}
\end{equation}
where ``$\cdots$'' denotes ``friction terms'' [e.g. $E \dot{\zeta}$] or ``mass terms'' (e.g. ``$\delta\varphi^2$'').
The kinetic terms are determined by the symmetric matrix $\bm{G}$, of which the entries are given explicitly in Appendix \ref{sec:coeffkin}.

For our purpose, we would like to examine, at the level of linear perturbations, how many scalar degrees of freedom are excited.
If only the $k$-essence term is present, i.e. $G^{(0)}_{,X}\neq 0$, $G^{(1)} = G^{(2)}= G^{(3)}=0$ (this is just the case considered in \cite{Gumrukcuoglu:2014xba,Mukohyama:2014rca} where $G^{(0)}\propto X^{(0)}$), by plugging the explicit expressions in Appendix \ref{sec:coeff} one may check, after some manipulations,
\begin{equation}
\det\left(\begin{array}{cc}
G_{22} & G_{23}\\
G_{23} & G_{33}
\end{array}\right)=0.
\end{equation}
This fact implies that between two linearly independent combinations of $\zeta$ and $\delta\varphi$, only one is dynamical, which can be given explicitly by\footnote{Note (\ref{Q_com}) takes the same form of the usual ``comoving curvature perturbation,'' which is a gauge invariant quantity in cosmological perturbations with a single scalar field.}
\begin{equation}
\zeta + \frac{G_{23}}{G_{22}} \delta\varphi \equiv \zeta - \frac{H}{\alpha^{(0)}\dot{\varphi}^{(0)}} \delta\varphi :=Q, \label{Q_com}
\end{equation}
where keep in mind that $\dot{\varphi}^{(0)}$ is defined in (\ref{dtddti}).
With this field redefinition, the two dynamical variables can be chosen to be $E$ and $Q$, while the remaining one becomes an auxiliary variable (and generating no further constraint).
In this simplest case, the system has two dynamical scalar modes as expected, one comes from the massive graviton, the other one comes from the Galileon field.

However, this property does not hold any more if we turn on other Horndeski/Galileon Lagrangians. 
For a nonvanishing $G^{(1)}$ with $G^{(0)} = G^{(2)} = G^{(3)} = 0$, we have
\begin{eqnarray}
\det \bm{G} & = & \frac{1}{C_{1}}36H^{2}k^{4}\, m_{BB}\big(\alpha^{(1)}\beta^{(1)}\dot{\bar{\phi}}\big)^{2} \nonumber\\
&  & \times \left[\big(a^{(1)}\big)^{2}X^{(1)}G_{,X}^{(1)}\big(a\dot{\bar{\phi}}+a_{0}\big)\right]^{2}, \label{detG_G1}
\end{eqnarray}
where $m_{BB}$ is given in (\ref{m_BB}). In (\ref{detG_G1}), $C_1$ is rather cumbersome and we prefer not to present its concrete form, which is irrelevant to the following analysis.
The form of (\ref{detG_G1}) implies that
\begin{equation}
\det \bm{G} \neq 0,\qquad \text{if}\quad m_{BB},\alpha^{(1)},\beta^{(1)}, G_{,X}^{(1)}\neq0.
\end{equation}
In the case of $G_{,X}^{(1)} \neq 0$ and $\beta^{(1)}\neq 0$, from (\ref{m_BB}), $m_{BB}$ acquires contributions proportional to $\beta^{(1)}$ and thus $m_{BB}\neq 0$ even in the absence of an explicit dRGT mass term.
In this case, a nonvanishing $\det \bm{G}$ implies all three scalar modes $E$, $\zeta$ and $\delta\varphi$ get excited.
In particular, besides the one come from the Galileon field, all 6 degrees of freedom in the spatial metric perturbations $H_{ij}$ become dynamical, and one of which may be ghost like.
In order to have a vanishing $\det \bm{G}$ and to reduce the number of dynamical degrees of freedom, one must have $G_{,X}^{(1)} = 0$ or $\alpha^{(1)}=0$ ($\beta^{(1)}=0$).
The former case is trivially equivalent to the $k$-essence case since $\mathcal{L}^{(1)}$ with $G^{(1)} = G^{(1)}(\varphi)$ is equivalent to $\mathcal{L}^{(0)}$ by integrating by parts, while from the above analysis, the $k$-essence with composite metric is always free of the BD ghost.
The later case implies the scalar field couples only to either the physical or fiducial metric, but never simultaneously.
To conclude, the Horndeski/Galileon term $\mathcal{L}^{(1)}$ generally reintroduces the BD ghost if the Galileon field coupled to the composite metric with $\alpha^{(1)}\neq 0$ and $\beta^{(1)} \neq 0$.

The same analysis can be applied to ``higher'' Horndeski/Galileon Lagrangians $\mathcal{L}^{(2)}$ and $\mathcal{L}^{(3)}$. For nonvanishing $G^{(2)}\neq 0$ with $G^{(0)}= G^{(1)} = G^{(3)}=0$, we have 
\begin{widetext}
\begin{eqnarray}
\det\bm{G} & = & \frac{1}{C_{2}}k^{4}\bar{N}^{3}a^{(2)}\left(\alpha^{(2)}\beta^{(2)}\dot{\bar{\phi}}\right)^{2}\Big[24a_{0}^{2}\big(\alpha^{(2)}\big)^{2}k^{2}\left(\beta^{(2)}\bar{N}\right)^{2}\big(a\dot{\bar{\phi}}-a_{0}\big)^{2}\left(G^{(2)}-2X^{(2)}G_{,X}^{(2)}\right)\nonumber \\
&  & +9m_{BB}a^{2}a^{(2)}\bar{N}\big(a\dot{\bar{\phi}}+a_{0}\big)^{2}\left(aN^{(2)}+2\big(\alpha^{(2)}\big)^{2}a^{(2)}\bar{N}\left(G^{(2)}-2X^{(2)}G_{,X}^{(2)}\right)\right)\Big], \label{detG_G2}
\end{eqnarray}
\end{widetext}
where $C_2$ is some factor of which the explicit expression is omitted here.
Again, (\ref{detG_G2}) is nonvanishing if $\alpha^{(2)}\neq 0$ and $\beta^{(2)}\neq 0$, which implies that there are three scalar modes are propagating.
Note $\mathcal{L}^{(2)}$ reduces to $R\big[g^{(2)}\big]$ when $G^{(2)} = 1$. 
The above fact thus prevents $R\big[g_{\mu\nu}(\alpha,\beta)\big]$ from being a viable ghost free derivative interaction term of the physical metric or between the physical and fiducial metric in bigravity theories\footnote{See however \cite{Noller:2014ioa} for a discussion on the derivative interactions among several metrics based on the composite metric. For other investigations on the non-GR derivative terms for gravitons, see \cite{Folkerts:2011ev,Hinterbichler:2013eza,Kimura:2013ika,deRham:2013tfa,Gao:2014jja}.}.

\section{Conclusion} \label{sec:con}

In this paper, we investigate the coupling between a Galileon scalar field and dRGT massive gravity through composite metrics proposed recently in \cite{deRham:2014naa}.
We present the full set of equations of motion for a spatially flat FRW background, and study tensor, vector and scalar perturbations around it.
If the scalar field is minimally coupled to the composite metric, i.e. only $\mathcal{L}^{(0)}$ in (\ref{LH0}) is present, we have shown explicitly that only two scalar modes are propagating at linear order. 
However, if other nonminimal/kinetic coupling terms (i.e. $\mathcal{L}^{(1)}$, $\mathcal{L}^{(2)}$ and $\mathcal{L}^{(3)}$) are present, generally the composite metric will excite all 6 degrees of freedom of the spatial metric perturbations, one of which may correspond to the BD ghost.
While it is important to investigate the mass of the would-be BD ghost mode, a detailed study is out of the scope of this paper.
It would also be interesting to explore possible ``doubly coupling'' to both the physical and fiducial metrics of the Galileon field without introducing the BD ghost.

% % % % % % % % % %
\acknowledgments

We would like to thank
Atsushi Naruko and 
Masahide Yamaguchi for collaboration at the initial stage and for valuable discussions and comments.
X.G. would like to thank the Institute of Astronomy and Space Science of Sun Yat-Sen University, the Institute of Theoretical Physics of Peking University, the Institute of High Energy Physics of Chinese Academy of Sciences and the Department of Physics of Beijing Normal University for their hospitality, during his visit in which part of this work was done.
X.G. also appreciates the elegant environment and the Grande Cappuccino of Zoo Coffee at Wudaokou, where many calculations of this work were done.
This work was in part supported by JSPS Grant-in-Aid for Scientific Research No. 25287054 (X.G.) and JSPS Research Fellowship for Young Scientists No. 26-11495 (D.Y.).

% % % % % % % % % % % % % % %
\appendix

% % % % % % % % % %
\begin{widetext}
\section{Coefficients in eq.(\ref{L2_s_ori})} \label{sec:coeff}

\begin{eqnarray}
m_{AA} & = & 1-\frac{1}{3H^{2}}\sum_{i=0}^{3}\alpha^{(i)}\frac{\bar{N}}{N^{(i)}}\rho^{(i)}-\frac{\big(\alpha^{(0)}\big)^{2}}{3H^{2}}\frac{\bar{N}\big(a^{(0)}\big)^{3}}{N^{(0)}a^{3}}\left(G^{(0)}+4\big(X^{(0)}\big)^{2}G_{,XX}^{(0)}\right)\nonumber \\
&  & +\frac{\big(\alpha^{(1)}\big)^{2}}{3H^{2}}\frac{\bar{N}\big(a^{(1)}\big)^{3}}{N^{(1)}a^{3}}\bigg[6\dot{\bar{\varphi}}^{(1)}H^{(1)}\left(2\big(X^{(1)}\big)^{2}G_{,XX}^{(1)}+3X^{(1)}G_{,X}^{(1)}\right)\nonumber \\
&  & \qquad\qquad\qquad-2\left(2\big(X^{(1)}\big)^{2}G_{,\tilde{X}\varphi}^{(1)}+X^{(1)}G_{,\varphi}^{(1)}\right)\bigg]\nonumber \\
&  & +\frac{\big(\alpha^{(2)}\big)^{2}}{3H^{2}}\frac{\bar{N}\big(a^{(2)}\big)^{3}}{N^{(2)}a^{3}}\bigg[6\big(H^{(2)}\big)^{2}\left(G^{(2)}-10X^{(2)}G_{,X}^{(2)}-28\big(X^{(2)}\big)^{2}G_{,XX}^{(2)}-8\big(X^{(2)}\big)^{3}G_{,XXX}^{(2)}\right)\nonumber \\
&  & \qquad\qquad\qquad+6H^{(2)}\dot{\bar{\varphi}}^{(2)}\left(G_{,\varphi}^{(2)}+8X^{(2)}G_{,X\varphi}^{(2)}+4\big(X^{(2)}\big)^{2}G_{,XX\varphi}^{(2)}\right)\bigg]\nonumber \\
&  & +\frac{\big(\alpha^{(3)}\big)^{2}}{3H^{2}}\frac{\bar{N}\big(a^{(3)}\big)^{3}}{N^{(3)}a^{3}}\bigg[6\big(H^{(3)}\big)^{2}\left(4\big(X^{(3)}\big)^{3}G_{,XX\varphi}^{\left(3\right)}+16\big(X^{(2)}\big)^{2}G_{,X\varphi}^{\left(3\right)}+9X^{(3)}G_{,\varphi}^{\left(3\right)}\right)\nonumber \\
&  & \qquad\qquad\qquad-2\big(H^{(3)}\big)^{3}\dot{\bar{\varphi}}^{(3)}\left(25X^{(3)}G_{,X}^{\left(3\right)}+24\big(X^{(3)}\big)^{2}G_{,XX}^{\left(3\right)}+4\big(X^{(3)}\big)^{3}G_{,XXX}^{\left(3\right)}\right)\bigg],\label{m_AA}
\end{eqnarray}
\begin{eqnarray}
g_{\varphi\varphi} & = & \frac{1}{2}\frac{\bar{N}\big(a^{(0)}\big)^{3}}{N^{(0)}a^{3}}\left(G_{,X}^{(0)}+2X^{(0)}G_{,XX}^{(0)}\right)\nonumber \\
&  & -\frac{\bar{N}\big(a^{(1)}\big)^{3}}{N^{(1)}a^{3}}\left[3H^{(1)}\dot{\bar{\varphi}}^{(1)}\left(G_{,X}^{(1)}+X^{(1)}G_{,XX}^{(1)}\right)-\left(X^{(1)}G_{,X\varphi}^{(1)}+G_{,\varphi}^{(1)}\right)\right]\nonumber \\
&  & +\frac{\bar{N}\big(a^{(2)}\big)^{3}}{N^{(2)}a^{3}}\bigg[3\big(H^{(2)}\big)^{2}\left(G_{,X}^{(2)}+8X^{(2)}G_{,XX}^{(2)}+4\big(X^{(2)}\big)^{2}G_{,XXX}^{(2)}\right)\nonumber \\
&  & \qquad\qquad\qquad-3H^{(2)}\dot{\bar{\varphi}}^{(2)}\left(2X^{(2)}G_{,XX\varphi}^{(2)}+3G_{,X\varphi}^{(2)}\right)\bigg]\nonumber \\
&  & +\frac{\bar{N}\big(a^{(3)}\big)^{3}}{N^{(3)}a^{3}}\bigg[\big(H^{(3)}\big)^{3}\dot{\bar{\varphi}}^{(3)}\left(3G_{,X}^{(3)}+7X^{(3)}G_{,XX}^{(3)}+2\big(X^{(2)}\big)^{2}G_{,XXX}^{(3)}\right)\nonumber \\
&  & \qquad\qquad\qquad-3\big(H^{(3)}\big)^{2}\left(2\big(X^{(2)}\big)^{2}G_{,XX\varphi}^{(3)}+5X^{(3)}G_{,X\varphi}^{(3)}+G_{,\varphi}^{(3)}\right)\bigg],\label{g_phiphi}
\end{eqnarray}
\begin{eqnarray}
w_{\varphi\varphi} & = & -\frac{1}{2}\frac{N^{(0)}a^{(0)}}{\bar{N}a}G_{,\tilde{X}}^{(0)}\nonumber \\
&  & +\frac{N^{(1)}a^{(1)}}{\bar{N}a}\bigg[-G_{,\varphi}^{(1)}+\ddot{\bar{\varphi}}^{(1)}X^{(1)}G_{,XX}^{(1)}+\left(\ddot{\bar{\varphi}}^{(1)}+2H^{(1)}\dot{\bar{\varphi}}^{(1)}\right)G_{,X}^{(1)}+X^{(1)}G_{,X\varphi}^{(1)}\bigg]\nonumber \\
&  & +\frac{N^{(2)}a^{(2)}}{\bar{N}a}\bigg[3\left(2H^{(2)}\dot{\bar{\varphi}}^{(2)}+\ddot{\bar{\varphi}}^{(2)}\right)G_{,X\varphi}^{(2)}+2X^{(2)}\left(\ddot{\bar{\varphi}}^{(2)}-2H^{(2)}\dot{\bar{\varphi}}^{(2)}\right)G_{,XX\varphi}^{(2)}\nonumber \\
&  & \qquad-\left(2\dot{H}^{(2)}+3\big(H^{(2)}\big)^{2}\right)G_{,X}^{(2)}+2X^{(2)}G_{,X\varphi\varphi}^{(2)}-4H^{(2)}X^{(2)}\dot{\bar{\varphi}}^{(2)}\ddot{\bar{\varphi}}^{(2)}G_{,XXX}^{(2)}\nonumber \\
&  & \qquad-\left(10\big(H^{(2)}\big)^{2}X^{(2)}+6H^{(2)}\dot{\bar{\varphi}}^{(2)}\ddot{\bar{\varphi}}^{(2)}+4X^{(2)}\dot{H}^{(2)}\right)G_{,XX}^{(2)}\bigg]\nonumber \\
&  & +\frac{N^{(3)}a^{(3)}}{\bar{N}a}\bigg[-\left(\big(H^{(3)}\big)^{2}\ddot{\bar{\varphi}}^{(3)}+\left(2\big(H^{(3)}\big)^{2}+2\dot{H}^{(3)}\right)H^{(3)}\dot{\bar{\varphi}}^{(3)}\right)G_{,X}^{(3)}\nonumber \\
&  & \qquad-\left(5\big(H^{(3)}\big)^{2}\ddot{\bar{\varphi}}^{(3)}+2\left(\big(H^{(3)}\big)^{2}+\dot{H}^{(3)}\right)H^{(3)}\dot{\bar{\varphi}}^{(3)}\right)X^{(3)}G_{,XX}^{(3)}\nonumber \\
&  & \qquad-2\big(H^{(3)}\big)^{2}\big(X^{(3)}\big)^{2}\ddot{\bar{\varphi}}^{(3)}G_{,XXX}^{(3)}-2H^{(3)}X^{(3)}\left(X^{(3)}H^{(3)}-\dot{\bar{\varphi}}^{(3)}\ddot{\bar{\varphi}}^{(3)}\right)G_{,XX\varphi}^{(3)}\nonumber \\
&  & \qquad+\left(2\dot{H}^{(3)}X^{(3)}+5\big(H^{(3)}\big)^{2}X^{(3)}+4H^{(3)}\dot{\bar{\varphi}}^{(3)}\ddot{\bar{\varphi}}^{(3)}\right)G_{,X\varphi}^{(3)}\nonumber \\
&  & \qquad+2H^{(3)}\dot{\bar{\varphi}}^{(3)}X^{(3)}G_{,X\varphi\varphi}^{(3)}+\left(2\dot{H}^{(3)}+3\big(H^{(3)}\big)^{2}\right)G_{,\varphi}^{(3)}\bigg],\label{w_phiphi}
\end{eqnarray}
\begin{eqnarray}
m_{\varphi\varphi} & = & \frac{1}{2}\frac{N^{(0)}\big(a^{(0)}\big)^{3}}{\bar{N}a^{3}}\left[-\left(\ddot{\bar{\varphi}}^{(0)}+3H^{(0)}\dot{\bar{\varphi}}^{(0)}\right)G_{,X\varphi}^{(0)}-\ddot{\bar{\varphi}}^{(0)}2X^{(0)}G_{,XX\varphi}^{(0)}-2X^{(0)}G_{,X\varphi\varphi}^{(0)}+G_{,\varphi\varphi}^{(0)}\right]\nonumber \\
&  & +\frac{N^{(1)}\big(a^{(1)}\big)^{3}}{\bar{N}a^{3}}\bigg[-X^{(1)}G_{,\varphi\varphi\varphi}^{(1)}+3H^{(1)}X^{(1)}G_{,XX\varphi}^{(1)}\dot{\bar{\varphi}}^{(1)}\ddot{\bar{\varphi}}^{(1)}\nonumber \\
&  & \qquad-G_{,\varphi\varphi}^{(1)}\left(\ddot{\bar{\varphi}}^{(1)}+3H^{(1)}\dot{\bar{\varphi}}^{(1)}\right)+X^{(1)}G_{,X\varphi\varphi}^{(1)}\left(3H^{(1)}\dot{\bar{\varphi}}^{(1)}-\ddot{\bar{\varphi}}^{(1)}\right)\nonumber \\
&  & \qquad+3G_{,X\varphi}^{(1)}\left(X^{(1)}\left(3\big(H^{(1)}\big)^{2}+\dot{H}^{(1)}\right)+H^{(1)}\dot{\bar{\varphi}}^{(1)}\ddot{\bar{\varphi}}^{(1)}\right)\bigg]\nonumber \\
&  & +\frac{N^{(2)}\big(a^{(2)}\big)^{3}}{\bar{N}a^{3}}\bigg[3\left(2X^{(2)}\dot{H}^{(2)}+6\big(H^{(2)}\big)^{2}X^{(2)}+3H^{(2)}\dot{\bar{\varphi}}^{(2)}\ddot{\bar{\varphi}}^{(2)}\right)G_{X\varphi\varphi}^{(2)}\nonumber \\
&  & \qquad+6X^{(2)}H^{(2)}\dot{\bar{\varphi}}^{(2)}G_{X\varphi\varphi\varphi}^{(2)}-6H^{(2)}X^{(2)}\left(2X^{(2)}H^{(2)}-\dot{\bar{\varphi}}^{(2)}\ddot{\bar{\varphi}}^{(2)}\right)G_{XX\varphi\varphi}^{(2)}\nonumber \\
&  & \qquad+3\left(2\big(H^{(2)}\big)^{2}+\dot{H}^{(2)}\right)G_{,\varphi\varphi}^{(2)}-12\big(H^{(2)}\big)^{2}\ddot{\bar{\varphi}}^{(2)}\big(X^{(2)}\big)^{2}G_{,XXX\varphi}^{(2)}\nonumber \\
&  & \qquad-6H^{(2)}X^{(2)}\left(3\big(H^{(2)}\big)^{2}\dot{\bar{\varphi}}^{(2)}+4H^{(2)}\ddot{\bar{\varphi}}^{(2)}+2\dot{\bar{\varphi}}^{(2)}\dot{H}^{(2)}\right)G_{,XX\varphi}^{(2)}\nonumber \\
&  & \qquad-3\left(\left(3\big(H^{(2)}\big)^{2}+2\dot{H}^{(2)}\right)H^{(2)}\dot{\bar{\varphi}}^{(2)}+\big(H^{(2)}\big)^{2}\ddot{\bar{\varphi}}^{(2)}\right)G_{,X\varphi}^{(2)}\bigg]\nonumber \\
&  & +\frac{N^{(3)}\big(a^{(3)}\big)^{3}}{\bar{N}a^{3}}\bigg[-2\big(H^{(3)}\big)^{2}\left(\dot{\bar{\varphi}}^{(3)}H^{(3)}-3\ddot{\bar{\varphi}}^{(3)}\right)\big(X^{(3)}\big)^{2}G_{,XX\varphi\varphi}^{(3)}\nonumber \\
&  & \qquad+H^{(3)}X^{(3)}\left(6\dot{\bar{\varphi}}^{(3)}\dot{H}^{(3)}+7\dot{\bar{\varphi}}^{(3)}\big(H^{(3)}\big)^{2}+15H^{(3)}\ddot{\bar{\varphi}}^{(3)}\right)G_{,X\varphi\varphi}^{(3)}\nonumber \\
&  & \qquad-\big(H^{(3)}\big)^{2}X^{(3)}\left(6\big(H^{(3)}\big)^{2}X^{(3)}+6X^{(3)}\dot{H}^{(3)}+7H^{(3)}\dot{\bar{\varphi}}^{(3)}\ddot{\bar{\varphi}}^{(3)}\right)G_{,XX\varphi}^{(3)}\nonumber \\
&  & \qquad+3H^{(3)}\left(\left(3\big(H^{(3)}\big)^{2}+2\dot{H}^{(3)}\right)\dot{\bar{\varphi}}^{(3)}+H^{(3)}\ddot{\bar{\varphi}}^{(3)}\right)G_{,\varphi\varphi}^{(3)}+3\big(H^{(3)}\big)^{2}X^{(3)}G_{,\varphi\varphi\varphi}^{(3)}\nonumber \\
&  & \qquad-2\big(H^{(3)}\big)^{3}\big(X^{(3)}\big)^{2}\dot{\bar{\varphi}}^{(3)}\ddot{\bar{\varphi}}^{(3)}G_{,XXX\varphi}^{(3)}+6\big(H^{(3)}\big)^{2}\big(X^{(3)}\big)^{2}G_{,X\varphi\varphi\varphi}^{(3)}\nonumber \\
&  & \qquad-3\big(H^{(3)}\big)^{2}\left(3X^{(3)}\big(H^{(3)}\big)^{2}+3X^{(3)}\dot{H}^{(3)}+H^{(3)}\dot{\bar{\varphi}}^{(3)}\ddot{\bar{\varphi}}^{(3)}\right)G_{,X\varphi}^{(3)}\bigg],\label{m_phiphi}
\end{eqnarray}
\begin{equation}
w_{AB}=\sum_{i=1}^{3}\frac{b}{b^{(i)}}\frac{1+b^{(i)}}{1+b}f_{A\zeta}^{(i)},\label{w_AB}
\end{equation}
where $f_{A\zeta}^{(i)}$ are given in (\ref{f_Az1})-(\ref{f_Az3}),
\begin{equation}
w_{AE}=\sum_{i=2}^{3}\frac{b}{b^{(i)}}g_{\gamma\gamma}^{(i)},\label{w_AE}
\end{equation}
\begin{equation}
f_{A\zeta}=f_{A\zeta}^{(1)}+f_{A\zeta}^{(2)}+f_{A\zeta}^{(3)},\label{f_Az}
\end{equation}
with
\begin{eqnarray}
f_{A\zeta}^{(1)} & = & \big(\alpha^{(1)}\big)^{3}\frac{\bar{N}^{2}a^{(1)}}{\big(N^{(1)}\big)^{2}a}\frac{\dot{\bar{\varphi}}^{(1)}}{H^{(1)}}X^{(1)}G_{,X}^{(1)},\label{f_Az1}\\
f_{A\zeta}^{(2)} & = & \big(\alpha^{(2)}\big)^{3}\frac{\bar{N}^{2}a^{(2)}}{\big(N^{(2)}\big)^{2}a}\bigg[2\left(G^{(2)}-4X^{(2)}G_{,X}^{(2)}-4\big(X^{(2)}\big)^{2}G_{,XX}^{(2)}\right)\nonumber \\
&  & \qquad\qquad\qquad+\frac{\dot{\bar{\varphi}}^{(2)}}{H^{(2)}}\left(2X^{(2)}G_{,X\varphi}^{(2)}+G_{,\varphi}^{(2)}\right)\bigg],\label{f_Az2}\\
f_{A\zeta}^{(3)} & = & \big(\alpha^{(3)}\big)^{3}\frac{\bar{N}^{2}a^{(3)}}{\big(N^{(3)}\big)^{2}a}\bigg[-H^{(3)}\dot{\bar{\varphi}}^{(3)}\left(5X^{(3)}G_{,X}^{\left(3\right)}+2\big(X^{(3)}\big)^{2}G_{,XX}^{\left(3\right)}\right)\nonumber \\
&  & \qquad\qquad\qquad+2\left(2\big(X^{(3)}\big)^{2}G_{,X\varphi}^{\left(3\right)}+3X^{(3)}G_{,\varphi}^{\left(3\right)}\right)\bigg],\label{f_Az3}
\end{eqnarray}
\begin{equation}
w_{A\zeta}=\sum_{i=2}^{3}\frac{b}{b^{(i)}}g_{\gamma\gamma}^{(i)},\label{w_Az}
\end{equation}
\begin{eqnarray}
f_{A\varphi} & = & -\alpha^{(0)}\frac{\bar{N}}{N^{(0)}}\left(\frac{a^{(0)}}{a}\right)^{3}\dot{\bar{\varphi}}^{(0)}\left(G_{,X}^{(0)}+2X^{(0)}G_{,XX}^{(0)}\right)\nonumber \\
&  & +\alpha^{(1)}\frac{\bar{N}}{N^{(1)}}\left(\frac{a^{(1)}}{a}\right)^{3}\bigg[6H^{(1)}X^{(1)}\left(3G_{,X}^{(1)}+2X^{(1)}G_{,XX}^{(1)}\right)-2\dot{\bar{\varphi}}^{(1)}\left(X^{(1)}G_{,X\varphi}^{(1)}+G_{,\varphi}^{(1)}\right)\bigg]\nonumber \\
&  & +\alpha^{(2)}\frac{\bar{N}}{N^{(2)}}\left(\frac{a^{(2)}}{a}\right)^{3}\bigg[-6\big(H^{(2)}\big)^{2}\dot{\bar{\varphi}}^{(2)}\left(3G_{,X}^{(2)}+12X^{(2)}G_{,XX}^{(2)}+4\big(X^{(2)}\big)^{2}G_{,XXX}^{(2)}\right)\nonumber \\
&  & \qquad+6H^{(2)}\left(4\big(X^{(2)}\big)^{2}G_{,XX\varphi}^{(2)}+8X^{(2)}G_{,X\varphi}^{(2)}+G_{,\varphi}^{(2)}\right)\bigg]\nonumber \\
&  & +\alpha^{(3)}\frac{\bar{N}}{N^{(3)}}\left(\frac{a^{(3)}}{a}\right)^{3}\bigg[-2\big(H^{(3)}\big)^{3}\left(15X^{(3)}G_{,X}^{(3)}+20\big(X^{(3)}\big)^{2}G_{,XX}^{(3)}+4\big(X^{(3)}\big)^{3}G_{,XXX}^{(3)}\right)\nonumber \\
&  & \qquad+6\big(H^{(3)}\big)^{2}\dot{\bar{\varphi}}^{(3)}\left(2\big(X^{(3)}\big)^{2}G_{,XX\varphi}^{(3)}+7X^{(3)}G_{,X\varphi}^{(3)}+3G_{,\varphi}^{(3)}\right)\bigg],\label{f_Aphi}
\end{eqnarray}
\end{widetext}
\begin{equation}
w_{A\varphi}=-\frac{1}{3}\sum_{i=1}^{3}\frac{b}{b^{(i)}}g_{\zeta\varphi}^{(i)},\label{w_Aphi}
\end{equation}
where $g_{\zeta\varphi}^{(i)}$ are given in (\ref{g_zphi1})-(\ref{g_zphi3}),
\begin{equation}
f_{BE}=\sum_{i=2}^{3}\frac{1+b^{(i)}}{1+b}\frac{b}{b^{(i)}}g_{\gamma\gamma}^{(i)},\label{f_BE}
\end{equation}
\begin{equation}
w_{BE}=-\frac{1}{3}H\sum_{i=2}^{3}\beta^{(i)}\frac{a_{0}}{a^{(i)}}\frac{1+b^{(i)}}{1+b}\frac{b}{b^{(i)}}g_{\gamma\gamma}^{(i)},\label{w_BE}
\end{equation}
\begin{equation}
f_{B\zeta}=\sum_{i=2}^{3}\frac{1+b^{(i)}}{1+b}\frac{b}{b^{(i)}}g_{\gamma\gamma}^{(i)},\label{f_Bz}
\end{equation}
\begin{equation}
w_{B\zeta}=-2H\sum_{i=2}^{3}\beta^{(i)}\frac{a_{0}}{a^{(i)}}\frac{1+b^{(i)}}{1+b}\frac{b}{b^{(i)}}g_{\gamma\gamma}^{(i)},\label{w_Bz}
\end{equation}
\begin{equation}
f_{B\varphi}\equiv\frac{1}{3}\sum_{i=1}^{3}\frac{b}{b^{(i)}}\frac{1+b^{(i)}}{1+b}g_{\zeta\varphi}^{(i)},\label{f_Bphi}
\end{equation}
\begin{equation}
w_{B\varphi}=-\sum_{i=0}^{3}\alpha^{(i)}\frac{a^{2}}{\big(a^{(i)}\big)^{2}}\frac{1+b^{(i)}}{1+b}\Gamma^{(i)},\label{w_Bphi}
\end{equation}
with
\begin{equation}
\Gamma^{(0)}  \equiv  \frac{N^{(0)}\big(a^{(0)}\big)^{3}}{\bar{N}a^{3}}\dot{\bar{\varphi}}^{(0)}G_{,X}^{(0)},\label{Gamma0}
\end{equation}
\begin{equation}
\Gamma^{(1)}  =  2\frac{N^{(1)}\big(a^{(1)}\big)^{3}}{\bar{N}a^{3}}\left(\dot{\bar{\varphi}}^{(1)}G_{,\varphi}^{(1)}-3H^{(1)}X^{(1)}G_{,X}^{(1)}\right),\label{Gamma1}
\end{equation}
\begin{align}
\Gamma^{(2)}  = & \frac{N^{(2)}\big(a^{(2)}\big)^{3}}{\bar{N}a^{3}}\bigg[-2H^{(2)}\left(10X^{(2)}G_{,X\varphi}^{(2)}+G_{,\varphi}^{(2)}\right)\nonumber \\
&   +6\big(H^{(2)}\big)^{2}\dot{\bar{\varphi}}^{(2)}\left(G_{,X}^{(2)}+2X^{(2)}G_{,XX}^{(2)}\right)+2\dot{\bar{\varphi}}^{(2)}G_{,\varphi\varphi}^{(2)}\bigg],\label{Gamma2}
\end{align}
\begin{align}
\Gamma^{(3)}  = & \frac{N^{(3)}\big(a^{(3)}\big)^{3}}{\bar{N}a^{3}}\bigg[2\big(H^{(3)}\big)^{3}X^{(3)}\left(3G_{,X}^{(3)}+2X^{(3)}G_{,XX}^{(3)}\right) \nonumber \\
& +4H^{(3)}X^{(3)}G_{,\varphi\varphi}^{(3)}  \nonumber\\
& -2\big(H^{(3)}\big)^{2}\dot{\bar{\varphi}}^{(3)}\left(4X^{(3)}G_{,X\varphi}^{(3)}+3G_{,\varphi}^{(3)}\right)\bigg],\label{Gamma3}
\end{align}
\begin{equation}
g_{\zeta\varphi}=g_{\zeta\varphi}^{(1)}+g_{\zeta\varphi}^{(2)}+g_{\zeta\varphi}^{(3)},\label{g_zphi}
\end{equation}
with
\begin{equation}
g_{\zeta\varphi}^{(1)}  =  -6\alpha^{(1)}\frac{\bar{N}\big(a^{(1)}\big)^{2}}{N^{(1)}a^{2}}X^{(1)}G_{,\tilde{X}}^{(1)},\label{g_zphi1}
\end{equation}
\begin{align}
g_{\zeta\varphi}^{(2)}  = & -6\alpha^{(2)}\frac{\bar{N}\big(a^{(2)}\big)^{2}}{N^{(2)}a^{2}}\Big[2X^{(2)}G_{,X\varphi}^{(2)}+G_{,\varphi}^{(2)} \nonumber\\
& -2H^{(2)}\dot{\bar{\varphi}}^{(2)}\left(G_{,X}^{(2)}+2X^{(2)}G_{,XX}^{(2)}\right)\Big],\label{g_zphi2}
\end{align}
\begin{align}
g_{\zeta\varphi}^{(3)} = & -6\alpha^{(3)}\frac{\bar{N}\big(a^{(3)}\big)^{2}}{N^{(3)}a^{2}}\bigg[2H^{(3)}\dot{\bar{\varphi}}^{(3)}\left(X^{(3)}G_{,X\varphi}^{(3)}+G_{,\varphi}^{(3)}\right)\nonumber \\
&  -\big(H^{(3)}\big)^{2}\left(3X^{(3)}G_{,X}^{(3)}+2\big(X^{(3)}\big)^{2}G_{,XX}^{(3)}\right)\bigg],\label{g_zphi3}
\end{align}
\begin{equation}
f_{\zeta\varphi}=\sum_{i=0}^{3}\frac{\bar{N}}{N^{(i)}}\left(3-a_{0}\frac{\partial}{\partial a_{0}}\right)\Gamma^{(i)}-2H\sum_{i=2}^{3}\frac{\beta^{(i)}}{\alpha^{(i)}}\frac{a_{0}}{a^{(i)}}\frac{\partial g_{\gamma\gamma}^{(i)}}{\partial\varphi},\label{f_zphi}
\end{equation}
where $\Gamma^{(i)}$ are given in (\ref{Gamma0})-(\ref{Gamma3}),
\begin{widetext}
\begin{eqnarray}
w_{\zeta\varphi} & = & 4\alpha^{(2)}\frac{N^{(2)}}{\bar{N}}\bigg[-2X^{(2)}G_{,X\varphi}^{(2)}+G_{,\varphi}^{(2)}-2\ddot{\bar{\varphi}}^{(2)}X^{(2)}G_{,XX}^{(2)}-G_{,X}^{(2)}\left(\ddot{\bar{\varphi}}^{(2)}+H^{(2)}\dot{\bar{\varphi}}^{(2)}\right)\bigg]\nonumber \\
&  & +4\alpha^{(3)}\frac{N^{(3)}}{\bar{N}}\bigg[X^{(3)}G_{,\varphi\varphi}^{(3)}-H^{(3)}\dot{\bar{\varphi}}^{(3)}\ddot{\bar{\varphi}}^{(3)}X^{(3)}G_{,XX}^{(3)}-\frac{1}{N^{(3)}a^{(3)}}\frac{\mathrm{d}}{\mathrm{d}t}\left(a^{(3)}H^{(3)}X^{(3)}\right)G_{,X}^{(3)}\nonumber \\
&  & \qquad\qquad+\left(\ddot{\bar{\varphi}}^{(3)}+H^{(3)}\dot{\bar{\varphi}}^{(3)}\right)G_{,\varphi}^{(3)}+\left(\frac{1}{2}\dot{\bar{\varphi}}^{(3)}\ddot{\bar{\varphi}}^{(3)}-H^{(3)}X^{(3)}\right)\dot{\bar{\varphi}}^{(3)}G_{,X\varphi}^{(3)}\bigg],\label{w_zphi}
\end{eqnarray}

\section{Coefficients of kinetic terms in eq.(\ref{L2s_kin})} \label{sec:coeffkin}

The entries of $\bm{G}$ in (\ref{L2s_kin}) are given explicitly by
\begin{eqnarray}
G_{11} & = & \frac{k^{4}}{12\Xi}\Big\{3a^{2}(1+g_{\gamma\gamma})(1+m_{AA})m_{BB}\nonumber \\
&  & -4k^{2}\left[f_{BE}\left(2+f_{BE}\right)(1+m_{AA})-2\left(1+g_{\gamma\gamma}\right)\left(1+w_{AB}\right)^{2}+m_{AA}+1\right]\Big\},\label{G11}\\
G_{12} & = & -\frac{2k^{4}}{\Xi}(1+f_{BE})\left[-2f_{A\ensuremath{\zeta}}(1+w_{AB})+f_{BE}m_{AA}+f_{BE}+m_{AA}-2w_{AB}-1\right],\label{G12}\\
G_{13} & = & \frac{k^{4}}{3H\Xi}(1+f_{BE})\left[3Hf_{B\ensuremath{\varphi}}(1+m_{AA})+2f_{A\ensuremath{\varphi}}(1+w_{AB})\right],\label{G13}\\
G_{22} & = & \frac{1}{\Xi}\Big\{9a^{2}m_{BB}\left[2f_{A\ensuremath{\zeta}}(2+f_{A\ensuremath{\zeta}})-g_{\gamma\gamma}(1+m_{AA})-m_{AA}+1\right]\nonumber \\
&  & +12k^{2}\Big[4f_{A\ensuremath{\zeta}}(1+f_{BE})(1+w_{AB})-f_{BE}^{2}(1+m_{AA})\nonumber \\
&  & +f_{BE}(-2m_{AA}+4w_{AB}+2)-2g_{\gamma\gamma}(1+w_{AB})^{2}-2w_{AB}^{2}-m_{\text{AA}}+1\Big]\Big\},\label{G22}\\
G_{23} & = & \frac{1}{2H\Xi}\Big\{3a^{2}m_{BB}(Hg_{\zeta\varphi}(m_{AA}+1)+2(f_{A\ensuremath{\zeta}}+1)f_{A\ensuremath{\varphi}})\nonumber \\
&  & +4k^{2}\Big[3Hf_{B\ensuremath{\varphi}}\big(-2(f_{A\ensuremath{\zeta}}+1)w_{AB}-2f_{A\ensuremath{\zeta}}+f_{BE}m_{AA}+f_{BE}+m_{AA}-1\big)\nonumber \\
&  & +2Hg_{\zeta\varphi}(w_{AB}+1){}^{2}+2f_{A\ensuremath{\varphi}}(f_{BE}+1)(w_{AB}+1)\Big]\Big\},\label{G23}\\
G_{33} & = & \frac{1}{2H^{2}\Xi}\Big\{ a^{2}m_{BB}\left(6H^{2}g_{\varphi\varphi}(m_{AA}+1)+f_{A\ensuremath{\varphi}}^{2}\right)\nonumber \\
&  & -2Hk^{2}\Big[H\left(3f_{B\ensuremath{\varphi}}^{2}(m_{AA}+1)-8g_{\varphi\varphi}(w_{AB}+1)^{2}\right)+4f_{A\ensuremath{\varphi}}f_{B\ensuremath{\varphi}}(w_{AB}+1)\Big]\Big\},\label{G33}
\end{eqnarray}
where various coefficients are given in Appendix \ref{sec:coeff}, and $\Xi$ is given in (\ref{Xi_def}).
\end{widetext}



\begin{thebibliography}{92}%
	\makeatletter
	\providecommand \@ifxundefined [1]{%
		\@ifx{#1\undefined}
	}%
	\providecommand \@ifnum [1]{%
		\ifnum #1\expandafter \@firstoftwo
		\else \expandafter \@secondoftwo
		\fi
	}%
	\providecommand \@ifx [1]{%
		\ifx #1\expandafter \@firstoftwo
		\else \expandafter \@secondoftwo
		\fi
	}%
	\providecommand \natexlab [1]{#1}%
	\providecommand \enquote  [1]{``#1''}%
	\providecommand \bibnamefont  [1]{#1}%
	\providecommand \bibfnamefont [1]{#1}%
	\providecommand \citenamefont [1]{#1}%
	\providecommand \href@noop [0]{\@secondoftwo}%
	\providecommand \href [0]{\begingroup \@sanitize@url \@href}%
	\providecommand \@href[1]{\@@startlink{#1}\@@href}%
	\providecommand \@@href[1]{\endgroup#1\@@endlink}%
	\providecommand \@sanitize@url [0]{\catcode `\\12\catcode `\$12\catcode
		`\&12\catcode `\#12\catcode `\^12\catcode `\_12\catcode `\%12\relax}%
	\providecommand \@@startlink[1]{}%
	\providecommand \@@endlink[0]{}%
	\providecommand \url  [0]{\begingroup\@sanitize@url \@url }%
	\providecommand \@url [1]{\endgroup\@href {#1}{\urlprefix }}%
	\providecommand \urlprefix  [0]{URL }%
	\providecommand \Eprint [0]{\href }%
	\providecommand \doibase [0]{http://dx.doi.org/}%
	\providecommand \selectlanguage [0]{\@gobble}%
	\providecommand \bibinfo  [0]{\@secondoftwo}%
	\providecommand \bibfield  [0]{\@secondoftwo}%
	\providecommand \translation [1]{[#1]}%
	\providecommand \BibitemOpen [0]{}%
	\providecommand \bibitemStop [0]{}%
	\providecommand \bibitemNoStop [0]{.\EOS\space}%
	\providecommand \EOS [0]{\spacefactor3000\relax}%
	\providecommand \BibitemShut  [1]{\csname bibitem#1\endcsname}%
	\let\auto@bib@innerbib\@empty
	%</preamble>
	\bibitem [{\citenamefont {Clifton}\ \emph {et~al.}(2012)\citenamefont
		{Clifton}, \citenamefont {Ferreira}, \citenamefont {Padilla},\ and\
		\citenamefont {Skordis}}]{Clifton:2011jh}%
	\BibitemOpen
	\bibfield  {author} {\bibinfo {author} {\bibfnamefont {T.}~\bibnamefont
			{Clifton}}, \bibinfo {author} {\bibfnamefont {P.~G.}\ \bibnamefont
			{Ferreira}}, \bibinfo {author} {\bibfnamefont {A.}~\bibnamefont {Padilla}}, \
		and\ \bibinfo {author} {\bibfnamefont {C.}~\bibnamefont {Skordis}},\ }\href
	{\doibase 10.1016/j.physrep.2012.01.001} {\bibfield  {journal} {\bibinfo
			{journal} {Phys.Rept.}\ }\textbf {\bibinfo {volume} {513}},\ \bibinfo {pages}
		{1} (\bibinfo {year} {2012})},\ \Eprint {http://arxiv.org/abs/1106.2476}
	{arXiv:1106.2476 [astro-ph.CO]} \BibitemShut {NoStop}%
	%%CITATION = ARXIV:1106.2476;%%
	\bibitem [{\citenamefont {Khoury}(2013)}]{Khoury:2013tda}%
	\BibitemOpen
	\bibfield  {author} {\bibinfo {author} {\bibfnamefont {J.}~\bibnamefont
			{Khoury}},\ }\href@noop {} {\  (\bibinfo {year} {2013})},\ \Eprint
	{http://arxiv.org/abs/1312.2006} {arXiv:1312.2006 [astro-ph.CO]} \BibitemShut
	{NoStop}%
	%%CITATION = ARXIV:1312.2006;%%
	\bibitem [{\citenamefont {Joyce}\ \emph {et~al.}(2015)\citenamefont {Joyce},
		\citenamefont {Jain}, \citenamefont {Khoury},\ and\ \citenamefont
		{Trodden}}]{Joyce:2014kja}%
	\BibitemOpen
	\bibfield  {author} {\bibinfo {author} {\bibfnamefont {A.}~\bibnamefont
			{Joyce}}, \bibinfo {author} {\bibfnamefont {B.}~\bibnamefont {Jain}},
		\bibinfo {author} {\bibfnamefont {J.}~\bibnamefont {Khoury}}, \ and\ \bibinfo
		{author} {\bibfnamefont {M.}~\bibnamefont {Trodden}},\ }\href {\doibase
		10.1016/j.physrep.2014.12.002} {\bibfield  {journal} {\bibinfo  {journal}
			{Phys.Rept.}\ }\textbf {\bibinfo {volume} {568}},\ \bibinfo {pages} {1}
		(\bibinfo {year} {2015})},\ \Eprint {http://arxiv.org/abs/1407.0059}
	{arXiv:1407.0059 [astro-ph.CO]} \BibitemShut {NoStop}%
	%%CITATION = ARXIV:1407.0059;%%
	\bibitem [{\citenamefont {Horndeski}(1974)}]{Horndeski:1974wa}%
	\BibitemOpen
	\bibfield  {author} {\bibinfo {author} {\bibfnamefont {G.~W.}\ \bibnamefont
			{Horndeski}},\ }\href {\doibase 10.1007/BF01807638} {\bibfield  {journal}
		{\bibinfo  {journal} {Int.J.Theor.Phys.}\ }\textbf {\bibinfo {volume} {10}},\
		\bibinfo {pages} {363} (\bibinfo {year} {1974})}\BibitemShut {NoStop}%
	%%CITATION = IJTPB,10,363;%%
	\bibitem [{\citenamefont {Deffayet}\ \emph {et~al.}(2011)\citenamefont
		{Deffayet}, \citenamefont {Gao}, \citenamefont {Steer},\ and\ \citenamefont
		{Zahariade}}]{Deffayet:2011gz}%
	\BibitemOpen
	\bibfield  {author} {\bibinfo {author} {\bibfnamefont {C.}~\bibnamefont
			{Deffayet}}, \bibinfo {author} {\bibfnamefont {X.}~\bibnamefont {Gao}},
		\bibinfo {author} {\bibfnamefont {D.}~\bibnamefont {Steer}}, \ and\ \bibinfo
		{author} {\bibfnamefont {G.}~\bibnamefont {Zahariade}},\ }\href {\doibase
		10.1103/PhysRevD.84.064039} {\bibfield  {journal} {\bibinfo  {journal}
			{Phys.Rev.}\ }\textbf {\bibinfo {volume} {D84}},\ \bibinfo {pages} {064039}
		(\bibinfo {year} {2011})},\ \Eprint {http://arxiv.org/abs/1103.3260}
	{arXiv:1103.3260 [hep-th]} \BibitemShut {NoStop}%
	%%CITATION = ARXIV:1103.3260;%%
	\bibitem [{\citenamefont {Deffayet}\ and\ \citenamefont
		{Steer}(2013)}]{Deffayet:2013lga}%
	\BibitemOpen
	\bibfield  {author} {\bibinfo {author} {\bibfnamefont {C.}~\bibnamefont
			{Deffayet}}\ and\ \bibinfo {author} {\bibfnamefont {D.~A.}\ \bibnamefont
			{Steer}},\ }\href {\doibase 10.1088/0264-9381/30/21/214006} {\bibfield
		{journal} {\bibinfo  {journal} {Class.Quant.Grav.}\ }\textbf {\bibinfo
			{volume} {30}},\ \bibinfo {pages} {214006} (\bibinfo {year} {2013})},\
	\Eprint {http://arxiv.org/abs/1307.2450} {arXiv:1307.2450 [hep-th]}
	\BibitemShut {NoStop}%
	%%CITATION = ARXIV:1307.2450;%%
	\bibitem [{\citenamefont {Charmousis}(2015)}]{Charmousis:2014mia}%
	\BibitemOpen
	\bibfield  {author} {\bibinfo {author} {\bibfnamefont {C.}~\bibnamefont
			{Charmousis}},\ }\bibfield  {booktitle} {\emph {\bibinfo {booktitle}
			{{Proceedings of the 7th Aegean Summer School : Beyond Einstein's theory of
					gravity. Modifications of Einstein's Theory of Gravity at Large
					Distances.}}},\ }\href {\doibase 10.1007/978-3-319-10070-8_2} {\bibfield
		{journal} {\bibinfo  {journal} {Lect. Notes Phys.}\ }\textbf {\bibinfo
			{volume} {892}},\ \bibinfo {pages} {25} (\bibinfo {year} {2015})},\ \Eprint
	{http://arxiv.org/abs/1405.1612} {arXiv:1405.1612 [gr-qc]} \BibitemShut
	{NoStop}%
	%%CITATION = ARXIV:1405.1612;%%
	\bibitem [{\citenamefont {Nicolis}\ \emph {et~al.}(2009)\citenamefont
		{Nicolis}, \citenamefont {Rattazzi},\ and\ \citenamefont
		{Trincherini}}]{Nicolis:2008in}%
	\BibitemOpen
	\bibfield  {author} {\bibinfo {author} {\bibfnamefont {A.}~\bibnamefont
			{Nicolis}}, \bibinfo {author} {\bibfnamefont {R.}~\bibnamefont {Rattazzi}}, \
		and\ \bibinfo {author} {\bibfnamefont {E.}~\bibnamefont {Trincherini}},\
	}\href {\doibase 10.1103/PhysRevD.79.064036} {\bibfield  {journal} {\bibinfo
		{journal} {Phys.Rev.}\ }\textbf {\bibinfo {volume} {D79}},\ \bibinfo {pages}
	{064036} (\bibinfo {year} {2009})},\ \Eprint {http://arxiv.org/abs/0811.2197}
{arXiv:0811.2197 [hep-th]} \BibitemShut {NoStop}%
%%CITATION = ARXIV:0811.2197;%%
\bibitem [{\citenamefont {Deffayet}\ \emph
	{et~al.}(2009{\natexlab{a}})\citenamefont {Deffayet}, \citenamefont
	{Esposito-Farese},\ and\ \citenamefont {Vikman}}]{Deffayet:2009wt}%
\BibitemOpen
\bibfield  {author} {\bibinfo {author} {\bibfnamefont {C.}~\bibnamefont
		{Deffayet}}, \bibinfo {author} {\bibfnamefont {G.}~\bibnamefont
		{Esposito-Farese}}, \ and\ \bibinfo {author} {\bibfnamefont {A.}~\bibnamefont
		{Vikman}},\ }\href {\doibase 10.1103/PhysRevD.79.084003} {\bibfield
	{journal} {\bibinfo  {journal} {Phys.Rev.}\ }\textbf {\bibinfo {volume}
		{D79}},\ \bibinfo {pages} {084003} (\bibinfo {year} {2009}{\natexlab{a}})},\
\Eprint {http://arxiv.org/abs/0901.1314} {arXiv:0901.1314 [hep-th]}
\BibitemShut {NoStop}%
%%CITATION = ARXIV:0901.1314;%%
\bibitem [{\citenamefont {Deffayet}\ \emph
	{et~al.}(2009{\natexlab{b}})\citenamefont {Deffayet}, \citenamefont {Deser},\
	and\ \citenamefont {Esposito-Farese}}]{Deffayet:2009mn}%
\BibitemOpen
\bibfield  {author} {\bibinfo {author} {\bibfnamefont {C.}~\bibnamefont
		{Deffayet}}, \bibinfo {author} {\bibfnamefont {S.}~\bibnamefont {Deser}}, \
	and\ \bibinfo {author} {\bibfnamefont {G.}~\bibnamefont {Esposito-Farese}},\
}\href {\doibase 10.1103/PhysRevD.80.064015} {\bibfield  {journal} {\bibinfo
	{journal} {Phys.Rev.}\ }\textbf {\bibinfo {volume} {D80}},\ \bibinfo {pages}
{064015} (\bibinfo {year} {2009}{\natexlab{b}})},\ \Eprint
{http://arxiv.org/abs/0906.1967} {arXiv:0906.1967 [gr-qc]} \BibitemShut
{NoStop}%
%%CITATION = ARXIV:0906.1967;%%
\bibitem [{\citenamefont {Kobayashi}\ \emph {et~al.}(2011)\citenamefont
	{Kobayashi}, \citenamefont {Yamaguchi},\ and\ \citenamefont
	{Yokoyama}}]{Kobayashi:2011nu}%
\BibitemOpen
\bibfield  {author} {\bibinfo {author} {\bibfnamefont {T.}~\bibnamefont
		{Kobayashi}}, \bibinfo {author} {\bibfnamefont {M.}~\bibnamefont
		{Yamaguchi}}, \ and\ \bibinfo {author} {\bibfnamefont {J.}~\bibnamefont
		{Yokoyama}},\ }\href {\doibase 10.1143/PTP.126.511} {\bibfield  {journal}
	{\bibinfo  {journal} {Prog.Theor.Phys.}\ }\textbf {\bibinfo {volume} {126}},\
	\bibinfo {pages} {511} (\bibinfo {year} {2011})},\ \Eprint
{http://arxiv.org/abs/1105.5723} {arXiv:1105.5723 [hep-th]} \BibitemShut
{NoStop}%
%%CITATION = ARXIV:1105.5723;%%
\bibitem [{\citenamefont {Fierz}\ and\ \citenamefont
	{Pauli}(1939)}]{Fierz:1939ix}%
\BibitemOpen
\bibfield  {author} {\bibinfo {author} {\bibfnamefont {M.}~\bibnamefont
		{Fierz}}\ and\ \bibinfo {author} {\bibfnamefont {W.}~\bibnamefont {Pauli}},\
}\href {\doibase 10.1098/rspa.1939.0140} {\bibfield  {journal} {\bibinfo
	{journal} {Proc.Roy.Soc.Lond.}\ }\textbf {\bibinfo {volume} {A173}},\
\bibinfo {pages} {211} (\bibinfo {year} {1939})}\BibitemShut {NoStop}%
%%CITATION = PRSLA,A173,211;%%
\bibitem [{\citenamefont {van Dam}\ and\ \citenamefont
	{Veltman}(1970)}]{vanDam:1970vg}%
\BibitemOpen
\bibfield  {author} {\bibinfo {author} {\bibfnamefont {H.}~\bibnamefont {van
			Dam}}\ and\ \bibinfo {author} {\bibfnamefont {M.}~\bibnamefont {Veltman}},\
}\href {\doibase 10.1016/0550-3213(70)90416-5} {\bibfield  {journal}
{\bibinfo  {journal} {Nucl.Phys.}\ }\textbf {\bibinfo {volume} {B22}},\
\bibinfo {pages} {397} (\bibinfo {year} {1970})}\BibitemShut {NoStop}%
%%CITATION = NUPHA,B22,397;%%
\bibitem [{\citenamefont {Zakharov}(1970)}]{Zakharov:1970cc}%
\BibitemOpen
\bibfield  {author} {\bibinfo {author} {\bibfnamefont {V.}~\bibnamefont
		{Zakharov}},\ }\href@noop {} {\bibfield  {journal} {\bibinfo  {journal} {JETP
			Lett.}\ }\textbf {\bibinfo {volume} {12}},\ \bibinfo {pages} {312} (\bibinfo
	{year} {1970})}\BibitemShut {NoStop}%
%%CITATION = JTPLA,12,312;%%
\bibitem [{\citenamefont {Vainshtein}(1972)}]{Vainshtein:1972sx}%
\BibitemOpen
\bibfield  {author} {\bibinfo {author} {\bibfnamefont {A.}~\bibnamefont
		{Vainshtein}},\ }\href {\doibase 10.1016/0370-2693(72)90147-5} {\bibfield
	{journal} {\bibinfo  {journal} {Phys.Lett.}\ }\textbf {\bibinfo {volume}
		{B39}},\ \bibinfo {pages} {393} (\bibinfo {year} {1972})}\BibitemShut
{NoStop}%
%%CITATION = PHLTA,B39,393;%%
\bibitem [{\citenamefont {Boulware}\ and\ \citenamefont
	{Deser}(1972)}]{Boulware:1973my}%
\BibitemOpen
\bibfield  {author} {\bibinfo {author} {\bibfnamefont {D.}~\bibnamefont
		{Boulware}}\ and\ \bibinfo {author} {\bibfnamefont {S.}~\bibnamefont
		{Deser}},\ }\href {\doibase 10.1103/PhysRevD.6.3368} {\bibfield  {journal}
	{\bibinfo  {journal} {Phys.Rev.}\ }\textbf {\bibinfo {volume} {D6}},\
	\bibinfo {pages} {3368} (\bibinfo {year} {1972})}\BibitemShut {NoStop}%
%%CITATION = PHRVA,D6,3368;%%
\bibitem [{\citenamefont {Arkani-Hamed}\ \emph {et~al.}(2003)\citenamefont
	{Arkani-Hamed}, \citenamefont {Georgi},\ and\ \citenamefont
	{Schwartz}}]{ArkaniHamed:2002sp}%
\BibitemOpen
\bibfield  {author} {\bibinfo {author} {\bibfnamefont {N.}~\bibnamefont
		{Arkani-Hamed}}, \bibinfo {author} {\bibfnamefont {H.}~\bibnamefont
		{Georgi}}, \ and\ \bibinfo {author} {\bibfnamefont {M.~D.}\ \bibnamefont
		{Schwartz}},\ }\href {\doibase 10.1016/S0003-4916(03)00068-X} {\bibfield
	{journal} {\bibinfo  {journal} {Annals Phys.}\ }\textbf {\bibinfo {volume}
		{305}},\ \bibinfo {pages} {96} (\bibinfo {year} {2003})},\ \Eprint
{http://arxiv.org/abs/hep-th/0210184} {arXiv:hep-th/0210184 [hep-th]}
\BibitemShut {NoStop}%
%%CITATION = HEP-TH/0210184;%%
\bibitem [{\citenamefont {de~Rham}\ and\ \citenamefont
	{Gabadadze}(2010)}]{deRham:2010ik}%
\BibitemOpen
\bibfield  {author} {\bibinfo {author} {\bibfnamefont {C.}~\bibnamefont
		{de~Rham}}\ and\ \bibinfo {author} {\bibfnamefont {G.}~\bibnamefont
		{Gabadadze}},\ }\href {\doibase 10.1103/PhysRevD.82.044020} {\bibfield
	{journal} {\bibinfo  {journal} {Phys.Rev.}\ }\textbf {\bibinfo {volume}
		{D82}},\ \bibinfo {pages} {044020} (\bibinfo {year} {2010})},\ \Eprint
{http://arxiv.org/abs/1007.0443} {arXiv:1007.0443 [hep-th]} \BibitemShut
{NoStop}%
%%CITATION = ARXIV:1007.0443;%%
\bibitem [{\citenamefont {de~Rham}\ \emph
	{et~al.}(2011{\natexlab{a}})\citenamefont {de~Rham}, \citenamefont
	{Gabadadze},\ and\ \citenamefont {Tolley}}]{deRham:2010kj}%
\BibitemOpen
\bibfield  {author} {\bibinfo {author} {\bibfnamefont {C.}~\bibnamefont
		{de~Rham}}, \bibinfo {author} {\bibfnamefont {G.}~\bibnamefont {Gabadadze}},
	\ and\ \bibinfo {author} {\bibfnamefont {A.~J.}\ \bibnamefont {Tolley}},\
}\href {\doibase 10.1103/PhysRevLett.106.231101} {\bibfield  {journal}
{\bibinfo  {journal} {Phys.Rev.Lett.}\ }\textbf {\bibinfo {volume} {106}},\
\bibinfo {pages} {231101} (\bibinfo {year} {2011}{\natexlab{a}})},\ \Eprint
{http://arxiv.org/abs/1011.1232} {arXiv:1011.1232 [hep-th]} \BibitemShut
{NoStop}%
%%CITATION = ARXIV:1011.1232;%%
\bibitem [{\citenamefont {de~Rham}(2014)}]{deRham:2014zqa}%
\BibitemOpen
\bibfield  {author} {\bibinfo {author} {\bibfnamefont {C.}~\bibnamefont
		{de~Rham}},\ }\href {\doibase 10.12942/lrr-2014-7} {\bibfield  {journal}
	{\bibinfo  {journal} {Living Rev.Rel.}\ }\textbf {\bibinfo {volume} {17}},\
	\bibinfo {pages} {7} (\bibinfo {year} {2014})},\ \Eprint
{http://arxiv.org/abs/1401.4173} {arXiv:1401.4173 [hep-th]} \BibitemShut
{NoStop}%
%%CITATION = ARXIV:1401.4173;%%
\bibitem [{\citenamefont {Hinterbichler}(2012)}]{Hinterbichler:2011tt}%
\BibitemOpen
\bibfield  {author} {\bibinfo {author} {\bibfnamefont {K.}~\bibnamefont
		{Hinterbichler}},\ }\href {\doibase 10.1103/RevModPhys.84.671} {\bibfield
	{journal} {\bibinfo  {journal} {Rev.Mod.Phys.}\ }\textbf {\bibinfo {volume}
		{84}},\ \bibinfo {pages} {671} (\bibinfo {year} {2012})},\ \Eprint
{http://arxiv.org/abs/1105.3735} {arXiv:1105.3735 [hep-th]} \BibitemShut
{NoStop}%
%%CITATION = ARXIV:1105.3735;%%
\bibitem [{\citenamefont {Rubakov}\ and\ \citenamefont
	{Tinyakov}(2008)}]{Rubakov:2008nh}%
\BibitemOpen
\bibfield  {author} {\bibinfo {author} {\bibfnamefont {V.}~\bibnamefont
		{Rubakov}}\ and\ \bibinfo {author} {\bibfnamefont {P.}~\bibnamefont
		{Tinyakov}},\ }\href {\doibase 10.1070/PU2008v051n08ABEH006600} {\bibfield
	{journal} {\bibinfo  {journal} {Phys.Usp.}\ }\textbf {\bibinfo {volume}
		{51}},\ \bibinfo {pages} {759} (\bibinfo {year} {2008})},\ \Eprint
{http://arxiv.org/abs/0802.4379} {arXiv:0802.4379 [hep-th]} \BibitemShut
{NoStop}%
%%CITATION = ARXIV:0802.4379;%%
\bibitem [{\citenamefont {Hassan}\ and\ \citenamefont
	{Rosen}(2012{\natexlab{a}})}]{Hassan:2011hr}%
\BibitemOpen
\bibfield  {author} {\bibinfo {author} {\bibfnamefont {S.}~\bibnamefont
		{Hassan}}\ and\ \bibinfo {author} {\bibfnamefont {R.~A.}\ \bibnamefont
		{Rosen}},\ }\href {\doibase 10.1103/PhysRevLett.108.041101} {\bibfield
	{journal} {\bibinfo  {journal} {Phys.Rev.Lett.}\ }\textbf {\bibinfo {volume}
		{108}},\ \bibinfo {pages} {041101} (\bibinfo {year} {2012}{\natexlab{a}})},\
\Eprint {http://arxiv.org/abs/1106.3344} {arXiv:1106.3344 [hep-th]}
\BibitemShut {NoStop}%
%%CITATION = ARXIV:1106.3344;%%
\bibitem [{\citenamefont {Hassan}\ \emph {et~al.}(2012)\citenamefont {Hassan},
	\citenamefont {Rosen},\ and\ \citenamefont {Schmidt-May}}]{Hassan:2011tf}%
\BibitemOpen
\bibfield  {author} {\bibinfo {author} {\bibfnamefont {S.}~\bibnamefont
		{Hassan}}, \bibinfo {author} {\bibfnamefont {R.~A.}\ \bibnamefont {Rosen}}, \
	and\ \bibinfo {author} {\bibfnamefont {A.}~\bibnamefont {Schmidt-May}},\
}\href {\doibase 10.1007/JHEP02(2012)026} {\bibfield  {journal} {\bibinfo
	{journal} {JHEP}\ }\textbf {\bibinfo {volume} {1202}},\ \bibinfo {pages}
{026} (\bibinfo {year} {2012})},\ \Eprint {http://arxiv.org/abs/1109.3230}
{arXiv:1109.3230 [hep-th]} \BibitemShut {NoStop}%
%%CITATION = ARXIV:1109.3230;%%
\bibitem [{\citenamefont {Deffayet}\ \emph {et~al.}(2013)\citenamefont
	{Deffayet}, \citenamefont {Mourad},\ and\ \citenamefont
	{Zahariade}}]{Deffayet:2012nr}%
\BibitemOpen
\bibfield  {author} {\bibinfo {author} {\bibfnamefont {C.}~\bibnamefont
		{Deffayet}}, \bibinfo {author} {\bibfnamefont {J.}~\bibnamefont {Mourad}}, \
	and\ \bibinfo {author} {\bibfnamefont {G.}~\bibnamefont {Zahariade}},\ }\href
{\doibase 10.1088/1475-7516/2013/01/032} {\bibfield  {journal} {\bibinfo
		{journal} {JCAP}\ }\textbf {\bibinfo {volume} {1301}},\ \bibinfo {pages}
	{032} (\bibinfo {year} {2013})},\ \Eprint {http://arxiv.org/abs/1207.6338}
{arXiv:1207.6338 [hep-th]} \BibitemShut {NoStop}%
%%CITATION = ARXIV:1207.6338;%%
\bibitem [{\citenamefont {Deser}\ and\ \citenamefont
	{Waldron}(2014)}]{Deser:2013rxa}%
\BibitemOpen
\bibfield  {author} {\bibinfo {author} {\bibfnamefont {S.}~\bibnamefont
		{Deser}}\ and\ \bibinfo {author} {\bibfnamefont {A.}~\bibnamefont
		{Waldron}},\ }\href {\doibase 10.1103/PhysRevD.89.027503} {\bibfield
	{journal} {\bibinfo  {journal} {Phys.Rev.}\ }\textbf {\bibinfo {volume}
		{D89}},\ \bibinfo {pages} {027503} (\bibinfo {year} {2014})},\ \Eprint
{http://arxiv.org/abs/1310.2675} {arXiv:1310.2675 [gr-qc]} \BibitemShut
{NoStop}%
%%CITATION = ARXIV:1310.2675;%%
\bibitem [{\citenamefont {Deser}\ \emph {et~al.}(2014)\citenamefont {Deser},
	\citenamefont {Sandora}, \citenamefont {Waldron},\ and\ \citenamefont
	{Zahariade}}]{Deser:2014hga}%
\BibitemOpen
\bibfield  {author} {\bibinfo {author} {\bibfnamefont {S.}~\bibnamefont
		{Deser}}, \bibinfo {author} {\bibfnamefont {M.}~\bibnamefont {Sandora}},
	\bibinfo {author} {\bibfnamefont {A.}~\bibnamefont {Waldron}}, \ and\
	\bibinfo {author} {\bibfnamefont {G.}~\bibnamefont {Zahariade}},\ }\href
{\doibase 10.1103/PhysRevD.90.104043} {\bibfield  {journal} {\bibinfo
		{journal} {Phys.Rev.}\ }\textbf {\bibinfo {volume} {D90}},\ \bibinfo {pages}
	{104043} (\bibinfo {year} {2014})},\ \Eprint {http://arxiv.org/abs/1408.0561}
{arXiv:1408.0561 [hep-th]} \BibitemShut {NoStop}%
%%CITATION = ARXIV:1408.0561;%%
\bibitem [{\citenamefont {Deser}\ \emph {et~al.}(2015)\citenamefont {Deser},
	\citenamefont {Izumi}, \citenamefont {Ong},\ and\ \citenamefont
	{Waldron}}]{Deser:2014fta}%
\BibitemOpen
\bibfield  {author} {\bibinfo {author} {\bibfnamefont {S.}~\bibnamefont
		{Deser}}, \bibinfo {author} {\bibfnamefont {K.}~\bibnamefont {Izumi}},
	\bibinfo {author} {\bibfnamefont {Y.~C.}\ \bibnamefont {Ong}}, \ and\
	\bibinfo {author} {\bibfnamefont {A.}~\bibnamefont {Waldron}},\ }\href
{\doibase 10.1142/S0217732315400064} {\bibfield  {journal} {\bibinfo
		{journal} {Mod. Phys. Lett.}\ }\textbf {\bibinfo {volume} {A30}},\ \bibinfo
	{pages} {1540006} (\bibinfo {year} {2015})},\ \Eprint
{http://arxiv.org/abs/1410.2289} {arXiv:1410.2289 [hep-th]} \BibitemShut
{NoStop}%
%%CITATION = ARXIV:1410.2289;%%
\bibitem [{\citenamefont {de~Rham}\ \emph
	{et~al.}(2011{\natexlab{b}})\citenamefont {de~Rham}, \citenamefont
	{Gabadadze}, \citenamefont {Heisenberg},\ and\ \citenamefont
	{Pirtskhalava}}]{deRham:2010tw}%
\BibitemOpen
\bibfield  {author} {\bibinfo {author} {\bibfnamefont {C.}~\bibnamefont
		{de~Rham}}, \bibinfo {author} {\bibfnamefont {G.}~\bibnamefont {Gabadadze}},
	\bibinfo {author} {\bibfnamefont {L.}~\bibnamefont {Heisenberg}}, \ and\
	\bibinfo {author} {\bibfnamefont {D.}~\bibnamefont {Pirtskhalava}},\ }\href
{\doibase 10.1103/PhysRevD.83.103516} {\bibfield  {journal} {\bibinfo
		{journal} {Phys.Rev.}\ }\textbf {\bibinfo {volume} {D83}},\ \bibinfo {pages}
	{103516} (\bibinfo {year} {2011}{\natexlab{b}})},\ \Eprint
{http://arxiv.org/abs/1010.1780} {arXiv:1010.1780 [hep-th]} \BibitemShut
{NoStop}%
%%CITATION = ARXIV:1010.1780;%%
\bibitem [{\citenamefont {D'Amico}\ \emph {et~al.}(2011)\citenamefont
	{D'Amico}, \citenamefont {de~Rham}, \citenamefont {Dubovsky}, \citenamefont
	{Gabadadze}, \citenamefont {Pirtskhalava} \emph {et~al.}}]{D'Amico:2011jj}%
\BibitemOpen
\bibfield  {author} {\bibinfo {author} {\bibfnamefont {G.}~\bibnamefont
		{D'Amico}}, \bibinfo {author} {\bibfnamefont {C.}~\bibnamefont {de~Rham}},
	\bibinfo {author} {\bibfnamefont {S.}~\bibnamefont {Dubovsky}}, \bibinfo
	{author} {\bibfnamefont {G.}~\bibnamefont {Gabadadze}}, \bibinfo {author}
	{\bibfnamefont {D.}~\bibnamefont {Pirtskhalava}},  \emph {et~al.},\ }\href
{\doibase 10.1103/PhysRevD.84.124046} {\bibfield  {journal} {\bibinfo
		{journal} {Phys.Rev.}\ }\textbf {\bibinfo {volume} {D84}},\ \bibinfo {pages}
	{124046} (\bibinfo {year} {2011})},\ \Eprint {http://arxiv.org/abs/1108.5231}
{arXiv:1108.5231 [hep-th]} \BibitemShut {NoStop}%
%%CITATION = ARXIV:1108.5231;%%
\bibitem [{\citenamefont {Gumrukcuoglu}\ \emph {et~al.}(2011)\citenamefont
	{Gumrukcuoglu}, \citenamefont {Lin},\ and\ \citenamefont
	{Mukohyama}}]{Gumrukcuoglu:2011ew}%
\BibitemOpen
\bibfield  {author} {\bibinfo {author} {\bibfnamefont {A.~E.}\ \bibnamefont
		{Gumrukcuoglu}}, \bibinfo {author} {\bibfnamefont {C.}~\bibnamefont {Lin}}, \
	and\ \bibinfo {author} {\bibfnamefont {S.}~\bibnamefont {Mukohyama}},\ }\href
{\doibase 10.1088/1475-7516/2011/11/030} {\bibfield  {journal} {\bibinfo
		{journal} {JCAP}\ }\textbf {\bibinfo {volume} {1111}},\ \bibinfo {pages}
	{030} (\bibinfo {year} {2011})},\ \Eprint {http://arxiv.org/abs/1109.3845}
{arXiv:1109.3845 [hep-th]} \BibitemShut {NoStop}%
%%CITATION = ARXIV:1109.3845;%%
\bibitem [{\citenamefont {Chamseddine}\ and\ \citenamefont
	{Volkov}(2011)}]{Chamseddine:2011bu}%
\BibitemOpen
\bibfield  {author} {\bibinfo {author} {\bibfnamefont {A.~H.}\ \bibnamefont
		{Chamseddine}}\ and\ \bibinfo {author} {\bibfnamefont {M.~S.}\ \bibnamefont
		{Volkov}},\ }\href {\doibase 10.1016/j.physletb.2011.09.085} {\bibfield
	{journal} {\bibinfo  {journal} {Phys.Lett.}\ }\textbf {\bibinfo {volume}
		{B704}},\ \bibinfo {pages} {652} (\bibinfo {year} {2011})},\ \Eprint
{http://arxiv.org/abs/1107.5504} {arXiv:1107.5504 [hep-th]} \BibitemShut
{NoStop}%
%%CITATION = ARXIV:1107.5504;%%
\bibitem [{\citenamefont {Gumrukcuoglu}\ \emph {et~al.}(2012)\citenamefont
	{Gumrukcuoglu}, \citenamefont {Lin},\ and\ \citenamefont
	{Mukohyama}}]{Gumrukcuoglu:2011zh}%
\BibitemOpen
\bibfield  {author} {\bibinfo {author} {\bibfnamefont {A.~E.}\ \bibnamefont
		{Gumrukcuoglu}}, \bibinfo {author} {\bibfnamefont {C.}~\bibnamefont {Lin}}, \
	and\ \bibinfo {author} {\bibfnamefont {S.}~\bibnamefont {Mukohyama}},\ }\href
{\doibase 10.1088/1475-7516/2012/03/006} {\bibfield  {journal} {\bibinfo
		{journal} {JCAP}\ }\textbf {\bibinfo {volume} {1203}},\ \bibinfo {pages}
	{006} (\bibinfo {year} {2012})},\ \Eprint {http://arxiv.org/abs/1111.4107}
{arXiv:1111.4107 [hep-th]} \BibitemShut {NoStop}%
%%CITATION = ARXIV:1111.4107;%%
\bibitem [{\citenamefont {Hassan}\ and\ \citenamefont
	{Rosen}(2011)}]{Hassan:2011vm}%
\BibitemOpen
\bibfield  {author} {\bibinfo {author} {\bibfnamefont {S.}~\bibnamefont
		{Hassan}}\ and\ \bibinfo {author} {\bibfnamefont {R.~A.}\ \bibnamefont
		{Rosen}},\ }\href {\doibase 10.1007/JHEP07(2011)009} {\bibfield  {journal}
	{\bibinfo  {journal} {JHEP}\ }\textbf {\bibinfo {volume} {1107}},\ \bibinfo
	{pages} {009} (\bibinfo {year} {2011})},\ \Eprint
{http://arxiv.org/abs/1103.6055} {arXiv:1103.6055 [hep-th]} \BibitemShut
{NoStop}%
%%CITATION = ARXIV:1103.6055;%%
\bibitem [{\citenamefont {Kobayashi}\ \emph {et~al.}(2012)\citenamefont
	{Kobayashi}, \citenamefont {Siino}, \citenamefont {Yamaguchi},\ and\
	\citenamefont {Yoshida}}]{Kobayashi:2012fz}%
\BibitemOpen
\bibfield  {author} {\bibinfo {author} {\bibfnamefont {T.}~\bibnamefont
		{Kobayashi}}, \bibinfo {author} {\bibfnamefont {M.}~\bibnamefont {Siino}},
	\bibinfo {author} {\bibfnamefont {M.}~\bibnamefont {Yamaguchi}}, \ and\
	\bibinfo {author} {\bibfnamefont {D.}~\bibnamefont {Yoshida}},\ }\href
{\doibase 10.1103/PhysRevD.86.061505} {\bibfield  {journal} {\bibinfo
		{journal} {Phys.Rev.}\ }\textbf {\bibinfo {volume} {D86}},\ \bibinfo {pages}
	{061505} (\bibinfo {year} {2012})},\ \Eprint {http://arxiv.org/abs/1205.4938}
{arXiv:1205.4938 [hep-th]} \BibitemShut {NoStop}%
%%CITATION = ARXIV:1205.4938;%%
\bibitem [{\citenamefont {de~Rham}\ and\ \citenamefont
	{Renaux-Petel}(2013)}]{deRham:2012kf}%
\BibitemOpen
\bibfield  {author} {\bibinfo {author} {\bibfnamefont {C.}~\bibnamefont
		{de~Rham}}\ and\ \bibinfo {author} {\bibfnamefont {S.}~\bibnamefont
		{Renaux-Petel}},\ }\href {\doibase 10.1088/1475-7516/2013/01/035} {\bibfield
	{journal} {\bibinfo  {journal} {JCAP}\ }\textbf {\bibinfo {volume} {1301}},\
	\bibinfo {pages} {035} (\bibinfo {year} {2013})},\ \Eprint
{http://arxiv.org/abs/1206.3482} {arXiv:1206.3482 [hep-th]} \BibitemShut
{NoStop}%
%%CITATION = ARXIV:1206.3482;%%
\bibitem [{\citenamefont {Fasiello}\ and\ \citenamefont
	{Tolley}(2012)}]{Fasiello:2012rw}%
\BibitemOpen
\bibfield  {author} {\bibinfo {author} {\bibfnamefont {M.}~\bibnamefont
		{Fasiello}}\ and\ \bibinfo {author} {\bibfnamefont {A.~J.}\ \bibnamefont
		{Tolley}},\ }\href {\doibase 10.1088/1475-7516/2012/11/035} {\bibfield
	{journal} {\bibinfo  {journal} {JCAP}\ }\textbf {\bibinfo {volume} {1211}},\
	\bibinfo {pages} {035} (\bibinfo {year} {2012})},\ \Eprint
{http://arxiv.org/abs/1206.3852} {arXiv:1206.3852 [hep-th]} \BibitemShut
{NoStop}%
%%CITATION = ARXIV:1206.3852;%%
\bibitem [{\citenamefont {Langlois}\ and\ \citenamefont
	{Naruko}(2012)}]{Langlois:2012hk}%
\BibitemOpen
\bibfield  {author} {\bibinfo {author} {\bibfnamefont {D.}~\bibnamefont
		{Langlois}}\ and\ \bibinfo {author} {\bibfnamefont {A.}~\bibnamefont
		{Naruko}},\ }\href {\doibase 10.1088/0264-9381/29/20/202001} {\bibfield
	{journal} {\bibinfo  {journal} {Class.Quant.Grav.}\ }\textbf {\bibinfo
		{volume} {29}},\ \bibinfo {pages} {202001} (\bibinfo {year} {2012})},\
\Eprint {http://arxiv.org/abs/1206.6810} {arXiv:1206.6810 [hep-th]}
\BibitemShut {NoStop}%
%%CITATION = ARXIV:1206.6810;%%
\bibitem [{\citenamefont {Langlois}\ and\ \citenamefont
	{Naruko}(2013)}]{Langlois:2013cya}%
\BibitemOpen
\bibfield  {author} {\bibinfo {author} {\bibfnamefont {D.}~\bibnamefont
		{Langlois}}\ and\ \bibinfo {author} {\bibfnamefont {A.}~\bibnamefont
		{Naruko}},\ }\href {\doibase 10.1088/0264-9381/30/20/205012} {\bibfield
	{journal} {\bibinfo  {journal} {Class.Quant.Grav.}\ }\textbf {\bibinfo
		{volume} {30}},\ \bibinfo {pages} {205012} (\bibinfo {year} {2013})},\
\Eprint {http://arxiv.org/abs/1305.6346} {arXiv:1305.6346 [hep-th]}
\BibitemShut {NoStop}%
%%CITATION = ARXIV:1305.6346;%%
\bibitem [{\citenamefont {Gao}\ \emph {et~al.}(2014)\citenamefont {Gao},
	\citenamefont {Kobayashi}, \citenamefont {Yamaguchi},\ and\ \citenamefont
	{Yoshida}}]{Gao:2014ula}%
\BibitemOpen
\bibfield  {author} {\bibinfo {author} {\bibfnamefont {X.}~\bibnamefont
		{Gao}}, \bibinfo {author} {\bibfnamefont {T.}~\bibnamefont {Kobayashi}},
	\bibinfo {author} {\bibfnamefont {M.}~\bibnamefont {Yamaguchi}}, \ and\
	\bibinfo {author} {\bibfnamefont {D.}~\bibnamefont {Yoshida}},\ }\href
{\doibase 10.1103/PhysRevD.90.124073} {\bibfield  {journal} {\bibinfo
		{journal} {Phys.Rev.}\ }\textbf {\bibinfo {volume} {D90}},\ \bibinfo {pages}
	{124073} (\bibinfo {year} {2014})},\ \Eprint {http://arxiv.org/abs/1409.3074}
{arXiv:1409.3074 [gr-qc]} \BibitemShut {NoStop}%
%%CITATION = ARXIV:1409.3074;%%
\bibitem [{\citenamefont {Kugo}\ and\ \citenamefont
	{Ohta}(2014)}]{Kugo:2014hja}%
\BibitemOpen
\bibfield  {author} {\bibinfo {author} {\bibfnamefont {T.}~\bibnamefont
		{Kugo}}\ and\ \bibinfo {author} {\bibfnamefont {N.}~\bibnamefont {Ohta}},\
}\href {\doibase 10.1093/ptep/ptu046} {\bibfield  {journal} {\bibinfo
	{journal} {PTEP}\ }\textbf {\bibinfo {volume} {2014}},\ \bibinfo {pages}
{043B04} (\bibinfo {year} {2014})},\ \Eprint {http://arxiv.org/abs/1401.3873}
{arXiv:1401.3873 [hep-th]} \BibitemShut {NoStop}%
%%CITATION = ARXIV:1401.3873;%%
\bibitem [{\citenamefont {De~Felice}\ \emph {et~al.}(2012)\citenamefont
	{De~Felice}, \citenamefont {Gumrukcuoglu},\ and\ \citenamefont
	{Mukohyama}}]{DeFelice:2012mx}%
\BibitemOpen
\bibfield  {author} {\bibinfo {author} {\bibfnamefont {A.}~\bibnamefont
		{De~Felice}}, \bibinfo {author} {\bibfnamefont {A.~E.}\ \bibnamefont
		{Gumrukcuoglu}}, \ and\ \bibinfo {author} {\bibfnamefont {S.}~\bibnamefont
		{Mukohyama}},\ }\href {\doibase 10.1103/PhysRevLett.109.171101} {\bibfield
	{journal} {\bibinfo  {journal} {Phys.Rev.Lett.}\ }\textbf {\bibinfo {volume}
		{109}},\ \bibinfo {pages} {171101} (\bibinfo {year} {2012})},\ \Eprint
{http://arxiv.org/abs/1206.2080} {arXiv:1206.2080 [hep-th]} \BibitemShut
{NoStop}%
%%CITATION = ARXIV:1206.2080;%%
\bibitem [{\citenamefont {Khosravi}\ \emph {et~al.}(2013)\citenamefont
	{Khosravi}, \citenamefont {Niz}, \citenamefont {Koyama},\ and\ \citenamefont
	{Tasinato}}]{Khosravi:2013axa}%
\BibitemOpen
\bibfield  {author} {\bibinfo {author} {\bibfnamefont {N.}~\bibnamefont
		{Khosravi}}, \bibinfo {author} {\bibfnamefont {G.}~\bibnamefont {Niz}},
	\bibinfo {author} {\bibfnamefont {K.}~\bibnamefont {Koyama}}, \ and\ \bibinfo
	{author} {\bibfnamefont {G.}~\bibnamefont {Tasinato}},\ }\href {\doibase
	10.1088/1475-7516/2013/08/044} {\bibfield  {journal} {\bibinfo  {journal}
		{JCAP}\ }\textbf {\bibinfo {volume} {1308}},\ \bibinfo {pages} {044}
	(\bibinfo {year} {2013})},\ \Eprint {http://arxiv.org/abs/1305.4950}
{arXiv:1305.4950 [hep-th]} \BibitemShut {NoStop}%
%%CITATION = ARXIV:1305.4950;%%
\bibitem [{\citenamefont {D'Amico}\ \emph {et~al.}(2013)\citenamefont
	{D'Amico}, \citenamefont {Gabadadze}, \citenamefont {Hui},\ and\
	\citenamefont {Pirtskhalava}}]{D'Amico:2012zv}%
\BibitemOpen
\bibfield  {author} {\bibinfo {author} {\bibfnamefont {G.}~\bibnamefont
		{D'Amico}}, \bibinfo {author} {\bibfnamefont {G.}~\bibnamefont {Gabadadze}},
	\bibinfo {author} {\bibfnamefont {L.}~\bibnamefont {Hui}}, \ and\ \bibinfo
	{author} {\bibfnamefont {D.}~\bibnamefont {Pirtskhalava}},\ }\href {\doibase
	10.1103/PhysRevD.87.064037} {\bibfield  {journal} {\bibinfo  {journal}
		{Phys.Rev.}\ }\textbf {\bibinfo {volume} {D87}},\ \bibinfo {pages} {064037}
	(\bibinfo {year} {2013})},\ \Eprint {http://arxiv.org/abs/1206.4253}
{arXiv:1206.4253 [hep-th]} \BibitemShut {NoStop}%
%%CITATION = ARXIV:1206.4253;%%
\bibitem [{\citenamefont {Huang}\ \emph {et~al.}(2012)\citenamefont {Huang},
	\citenamefont {Piao},\ and\ \citenamefont {Zhou}}]{Huang:2012pe}%
\BibitemOpen
\bibfield  {author} {\bibinfo {author} {\bibfnamefont {Q.-G.}\ \bibnamefont
		{Huang}}, \bibinfo {author} {\bibfnamefont {Y.-S.}\ \bibnamefont {Piao}}, \
	and\ \bibinfo {author} {\bibfnamefont {S.-Y.}\ \bibnamefont {Zhou}},\ }\href
{\doibase 10.1103/PhysRevD.86.124014} {\bibfield  {journal} {\bibinfo
		{journal} {Phys.Rev.}\ }\textbf {\bibinfo {volume} {D86}},\ \bibinfo {pages}
	{124014} (\bibinfo {year} {2012})},\ \Eprint {http://arxiv.org/abs/1206.5678}
{arXiv:1206.5678 [hep-th]} \BibitemShut {NoStop}%
%%CITATION = ARXIV:1206.5678;%%
\bibitem [{\citenamefont {De~Felice}\ and\ \citenamefont
	{Mukohyama}(2014)}]{DeFelice:2013tsa}%
\BibitemOpen
\bibfield  {author} {\bibinfo {author} {\bibfnamefont {A.}~\bibnamefont
		{De~Felice}}\ and\ \bibinfo {author} {\bibfnamefont {S.}~\bibnamefont
		{Mukohyama}},\ }\href {\doibase 10.1016/j.physletb.2013.12.041} {\bibfield
	{journal} {\bibinfo  {journal} {Phys.Lett.}\ }\textbf {\bibinfo {volume}
		{B728}},\ \bibinfo {pages} {622} (\bibinfo {year} {2014})},\ \Eprint
{http://arxiv.org/abs/1306.5502} {arXiv:1306.5502 [hep-th]} \BibitemShut
{NoStop}%
%%CITATION = ARXIV:1306.5502;%%
\bibitem [{\citenamefont {De~Felice}\ \emph
	{et~al.}(2013{\natexlab{a}})\citenamefont {De~Felice}, \citenamefont
	{Emir~Gümrükçüoğlu},\ and\ \citenamefont
	{Mukohyama}}]{DeFelice:2013dua}%
\BibitemOpen
\bibfield  {author} {\bibinfo {author} {\bibfnamefont {A.}~\bibnamefont
		{De~Felice}}, \bibinfo {author} {\bibfnamefont {A.}~\bibnamefont
		{Emir~Gümrükçüoğlu}}, \ and\ \bibinfo {author} {\bibfnamefont
		{S.}~\bibnamefont {Mukohyama}},\ }\href {\doibase 10.1103/PhysRevD.88.124006}
{\bibfield  {journal} {\bibinfo  {journal} {Phys.Rev.}\ }\textbf {\bibinfo
		{volume} {D88}},\ \bibinfo {pages} {124006} (\bibinfo {year}
	{2013}{\natexlab{a}})},\ \Eprint {http://arxiv.org/abs/1309.3162}
{arXiv:1309.3162 [hep-th]} \BibitemShut {NoStop}%
%%CITATION = ARXIV:1309.3162;%%
\bibitem [{\citenamefont {Gümrükçüoğlu}\ \emph {et~al.}(2013)\citenamefont
	{Gümrükçüoğlu}, \citenamefont {Hinterbichler}, \citenamefont {Lin},
	\citenamefont {Mukohyama},\ and\ \citenamefont
	{Trodden}}]{Gumrukcuoglu:2013nza}%
\BibitemOpen
\bibfield  {author} {\bibinfo {author} {\bibfnamefont {A.~E.}\ \bibnamefont
		{Gümrükçüoğlu}}, \bibinfo {author} {\bibfnamefont {K.}~\bibnamefont
		{Hinterbichler}}, \bibinfo {author} {\bibfnamefont {C.}~\bibnamefont {Lin}},
	\bibinfo {author} {\bibfnamefont {S.}~\bibnamefont {Mukohyama}}, \ and\
	\bibinfo {author} {\bibfnamefont {M.}~\bibnamefont {Trodden}},\ }\href
{\doibase 10.1103/PhysRevD.88.024023} {\bibfield  {journal} {\bibinfo
		{journal} {Phys.Rev.}\ }\textbf {\bibinfo {volume} {D88}},\ \bibinfo {pages}
	{024023} (\bibinfo {year} {2013})},\ \Eprint {http://arxiv.org/abs/1304.0449}
{arXiv:1304.0449 [hep-th]} \BibitemShut {NoStop}%
%%CITATION = ARXIV:1304.0449;%%
\bibitem [{\citenamefont {D’Amico}\ \emph {et~al.}(2013)\citenamefont
	{D’Amico}, \citenamefont {Gabadadze}, \citenamefont {Hui},\ and\
	\citenamefont {Pirtskhalava}}]{D'Amico:2013kya}%
\BibitemOpen
\bibfield  {author} {\bibinfo {author} {\bibfnamefont {G.}~\bibnamefont
		{D’Amico}}, \bibinfo {author} {\bibfnamefont {G.}~\bibnamefont
		{Gabadadze}}, \bibinfo {author} {\bibfnamefont {L.}~\bibnamefont {Hui}}, \
	and\ \bibinfo {author} {\bibfnamefont {D.}~\bibnamefont {Pirtskhalava}},\
}\href {\doibase 10.1088/0264-9381/30/18/184005} {\bibfield  {journal}
{\bibinfo  {journal} {Class.Quant.Grav.}\ }\textbf {\bibinfo {volume} {30}},\
\bibinfo {pages} {184005} (\bibinfo {year} {2013})},\ \Eprint
{http://arxiv.org/abs/1304.0723} {arXiv:1304.0723 [hep-th]} \BibitemShut
{NoStop}%
%%CITATION = ARXIV:1304.0723;%%
\bibitem [{\citenamefont {Gabadadze}\ \emph {et~al.}(2014)\citenamefont
	{Gabadadze}, \citenamefont {Kimura},\ and\ \citenamefont
	{Pirtskhalava}}]{Gabadadze:2014kaa}%
\BibitemOpen
\bibfield  {author} {\bibinfo {author} {\bibfnamefont {G.}~\bibnamefont
		{Gabadadze}}, \bibinfo {author} {\bibfnamefont {R.}~\bibnamefont {Kimura}}, \
	and\ \bibinfo {author} {\bibfnamefont {D.}~\bibnamefont {Pirtskhalava}},\
}\href {\doibase 10.1103/PhysRevD.90.024029} {\bibfield  {journal} {\bibinfo
	{journal} {Phys.Rev.}\ }\textbf {\bibinfo {volume} {D90}},\ \bibinfo {pages}
{024029} (\bibinfo {year} {2014})},\ \Eprint {http://arxiv.org/abs/1401.5403}
{arXiv:1401.5403 [hep-th]} \BibitemShut {NoStop}%
%%CITATION = ARXIV:1401.5403;%%
\bibitem [{\citenamefont {Kahniashvili}\ \emph {et~al.}(2015)\citenamefont
	{Kahniashvili}, \citenamefont {Kar}, \citenamefont {Lavrelashvili},
	\citenamefont {Agarwal}, \citenamefont {Heisenberg},\ and\ \citenamefont
	{Kosowsky}}]{Kahniashvili:2014wua}%
\BibitemOpen
\bibfield  {author} {\bibinfo {author} {\bibfnamefont {T.}~\bibnamefont
		{Kahniashvili}}, \bibinfo {author} {\bibfnamefont {A.}~\bibnamefont {Kar}},
	\bibinfo {author} {\bibfnamefont {G.}~\bibnamefont {Lavrelashvili}}, \bibinfo
	{author} {\bibfnamefont {N.}~\bibnamefont {Agarwal}}, \bibinfo {author}
	{\bibfnamefont {L.}~\bibnamefont {Heisenberg}}, \ and\ \bibinfo {author}
	{\bibfnamefont {A.}~\bibnamefont {Kosowsky}},\ }\href {\doibase
	10.1103/PhysRevD.91.041301} {\bibfield  {journal} {\bibinfo  {journal} {Phys.
			Rev.}\ }\textbf {\bibinfo {volume} {D91}},\ \bibinfo {pages} {041301}
	(\bibinfo {year} {2015})},\ \Eprint {http://arxiv.org/abs/1412.4300}
{arXiv:1412.4300 [astro-ph.CO]} \BibitemShut {NoStop}%
%%CITATION = ARXIV:1412.4300;%%
\bibitem [{\citenamefont {Nojiri}\ and\ \citenamefont
	{Odintsov}(2012)}]{Nojiri:2012zu}%
\BibitemOpen
\bibfield  {author} {\bibinfo {author} {\bibfnamefont {S.}~\bibnamefont
		{Nojiri}}\ and\ \bibinfo {author} {\bibfnamefont {S.~D.}\ \bibnamefont
		{Odintsov}},\ }\href {\doibase 10.1016/j.physletb.2012.08.049} {\bibfield
	{journal} {\bibinfo  {journal} {Phys.Lett.}\ }\textbf {\bibinfo {volume}
		{B716}},\ \bibinfo {pages} {377} (\bibinfo {year} {2012})},\ \Eprint
{http://arxiv.org/abs/1207.5106} {arXiv:1207.5106 [hep-th]} \BibitemShut
{NoStop}%
%%CITATION = ARXIV:1207.5106;%%
\bibitem [{\citenamefont {Klusoň}\ \emph {et~al.}(2013)\citenamefont
	{Klusoň}, \citenamefont {Nojiri},\ and\ \citenamefont
	{Odintsov}}]{Kluson:2013yaa}%
\BibitemOpen
\bibfield  {author} {\bibinfo {author} {\bibfnamefont {J.}~\bibnamefont
		{Klusoň}}, \bibinfo {author} {\bibfnamefont {S.}~\bibnamefont {Nojiri}}, \
	and\ \bibinfo {author} {\bibfnamefont {S.~D.}\ \bibnamefont {Odintsov}},\
}\href {\doibase 10.1016/j.physletb.2013.10.003} {\bibfield  {journal}
{\bibinfo  {journal} {Phys.Lett.}\ }\textbf {\bibinfo {volume} {B726}},\
\bibinfo {pages} {918} (\bibinfo {year} {2013})},\ \Eprint
{http://arxiv.org/abs/1309.2185} {arXiv:1309.2185 [hep-th]} \BibitemShut
{NoStop}%
%%CITATION = ARXIV:1309.2185;%%
\bibitem [{\citenamefont {Cai}\ \emph {et~al.}(2014)\citenamefont {Cai},
	\citenamefont {Duplessis},\ and\ \citenamefont {Saridakis}}]{Cai:2013lqa}%
\BibitemOpen
\bibfield  {author} {\bibinfo {author} {\bibfnamefont {Y.-F.}\ \bibnamefont
		{Cai}}, \bibinfo {author} {\bibfnamefont {F.}~\bibnamefont {Duplessis}}, \
	and\ \bibinfo {author} {\bibfnamefont {E.~N.}\ \bibnamefont {Saridakis}},\
}\href {\doibase 10.1103/PhysRevD.90.064051} {\bibfield  {journal} {\bibinfo
	{journal} {Phys.Rev.}\ }\textbf {\bibinfo {volume} {D90}},\ \bibinfo {pages}
{064051} (\bibinfo {year} {2014})},\ \Eprint {http://arxiv.org/abs/1307.7150}
{arXiv:1307.7150 [hep-th]} \BibitemShut {NoStop}%
%%CITATION = ARXIV:1307.7150;%%
\bibitem [{\citenamefont {Cai}\ and\ \citenamefont
	{Saridakis}(2014)}]{Cai:2014upa}%
\BibitemOpen
\bibfield  {author} {\bibinfo {author} {\bibfnamefont {Y.-F.}\ \bibnamefont
		{Cai}}\ and\ \bibinfo {author} {\bibfnamefont {E.~N.}\ \bibnamefont
		{Saridakis}},\ }\href {\doibase 10.1103/PhysRevD.90.063528} {\bibfield
	{journal} {\bibinfo  {journal} {Phys.Rev.}\ }\textbf {\bibinfo {volume}
		{D90}},\ \bibinfo {pages} {063528} (\bibinfo {year} {2014})},\ \Eprint
{http://arxiv.org/abs/1401.4418} {arXiv:1401.4418 [astro-ph.CO]} \BibitemShut
{NoStop}%
%%CITATION = ARXIV:1401.4418;%%
\bibitem [{\citenamefont {Wu}(2014)}]{Wu:2014hva}%
\BibitemOpen
\bibfield  {author} {\bibinfo {author} {\bibfnamefont {D.-J.}\ \bibnamefont
		{Wu}},\ }\href {\doibase 10.1103/PhysRevD.90.043530} {\bibfield  {journal}
	{\bibinfo  {journal} {Phys.Rev.}\ }\textbf {\bibinfo {volume} {D90}},\
	\bibinfo {pages} {043530} (\bibinfo {year} {2014})},\ \Eprint
{http://arxiv.org/abs/1403.4442} {arXiv:1403.4442 [hep-th]} \BibitemShut
{NoStop}%
%%CITATION = ARXIV:1403.4442;%%
\bibitem [{\citenamefont {De~Felice}\ \emph
	{et~al.}(2013{\natexlab{b}})\citenamefont {De~Felice}, \citenamefont
	{Gümrükçüoğlu}, \citenamefont {Lin},\ and\ \citenamefont
	{Mukohyama}}]{DeFelice:2013awa}%
\BibitemOpen
\bibfield  {author} {\bibinfo {author} {\bibfnamefont {A.}~\bibnamefont
		{De~Felice}}, \bibinfo {author} {\bibfnamefont {A.~E.}\ \bibnamefont
		{Gümrükçüoğlu}}, \bibinfo {author} {\bibfnamefont {C.}~\bibnamefont
		{Lin}}, \ and\ \bibinfo {author} {\bibfnamefont {S.}~\bibnamefont
		{Mukohyama}},\ }\href {\doibase 10.1088/1475-7516/2013/05/035} {\bibfield
	{journal} {\bibinfo  {journal} {JCAP}\ }\textbf {\bibinfo {volume} {1305}},\
	\bibinfo {pages} {035} (\bibinfo {year} {2013}{\natexlab{b}})},\ \Eprint
{http://arxiv.org/abs/1303.4154} {arXiv:1303.4154 [hep-th]} \BibitemShut
{NoStop}%
%%CITATION = ARXIV:1303.4154;%%
\bibitem [{\citenamefont {de~Rham}\ \emph
	{et~al.}(2014{\natexlab{a}})\citenamefont {de~Rham}, \citenamefont
	{Fasiello},\ and\ \citenamefont {Tolley}}]{deRham:2014gla}%
\BibitemOpen
\bibfield  {author} {\bibinfo {author} {\bibfnamefont {C.}~\bibnamefont
		{de~Rham}}, \bibinfo {author} {\bibfnamefont {M.}~\bibnamefont {Fasiello}}, \
	and\ \bibinfo {author} {\bibfnamefont {A.~J.}\ \bibnamefont {Tolley}},\
}\href {\doibase 10.1142/S0218271814430068} {\bibfield  {journal} {\bibinfo
	{journal} {Int. J. Mod. Phys.}\ }\textbf {\bibinfo {volume} {D23}},\ \bibinfo
{pages} {1443006} (\bibinfo {year} {2014}{\natexlab{a}})},\ \Eprint
{http://arxiv.org/abs/1410.0960} {arXiv:1410.0960 [hep-th]} \BibitemShut
{NoStop}%
%%CITATION = ARXIV:1410.0960;%%
\bibitem [{\citenamefont {Hassan}\ and\ \citenamefont
	{Rosen}(2012{\natexlab{b}})}]{Hassan:2011zd}%
\BibitemOpen
\bibfield  {author} {\bibinfo {author} {\bibfnamefont {S.}~\bibnamefont
		{Hassan}}\ and\ \bibinfo {author} {\bibfnamefont {R.~A.}\ \bibnamefont
		{Rosen}},\ }\href {\doibase 10.1007/JHEP02(2012)126} {\bibfield  {journal}
	{\bibinfo  {journal} {JHEP}\ }\textbf {\bibinfo {volume} {1202}},\ \bibinfo
	{pages} {126} (\bibinfo {year} {2012}{\natexlab{b}})},\ \Eprint
{http://arxiv.org/abs/1109.3515} {arXiv:1109.3515 [hep-th]} \BibitemShut
{NoStop}%
%%CITATION = ARXIV:1109.3515;%%
\bibitem [{\citenamefont {Akrami}\ \emph
	{et~al.}(2013{\natexlab{a}})\citenamefont {Akrami}, \citenamefont {Koivisto},
	\citenamefont {Mota},\ and\ \citenamefont {Sandstad}}]{Akrami:2013ffa}%
\BibitemOpen
\bibfield  {author} {\bibinfo {author} {\bibfnamefont {Y.}~\bibnamefont
		{Akrami}}, \bibinfo {author} {\bibfnamefont {T.~S.}\ \bibnamefont
		{Koivisto}}, \bibinfo {author} {\bibfnamefont {D.~F.}\ \bibnamefont {Mota}},
	\ and\ \bibinfo {author} {\bibfnamefont {M.}~\bibnamefont {Sandstad}},\
}\href {\doibase 10.1088/1475-7516/2013/10/046} {\bibfield  {journal}
{\bibinfo  {journal} {JCAP}\ }\textbf {\bibinfo {volume} {1310}},\ \bibinfo
{pages} {046} (\bibinfo {year} {2013}{\natexlab{a}})},\ \Eprint
{http://arxiv.org/abs/1306.0004} {arXiv:1306.0004 [hep-th]} \BibitemShut
{NoStop}%
%%CITATION = ARXIV:1306.0004;%%
\bibitem [{\citenamefont {Tamanini}\ \emph {et~al.}(2014)\citenamefont
	{Tamanini}, \citenamefont {Saridakis},\ and\ \citenamefont
	{Koivisto}}]{Tamanini:2013xia}%
\BibitemOpen
\bibfield  {author} {\bibinfo {author} {\bibfnamefont {N.}~\bibnamefont
		{Tamanini}}, \bibinfo {author} {\bibfnamefont {E.~N.}\ \bibnamefont
		{Saridakis}}, \ and\ \bibinfo {author} {\bibfnamefont {T.~S.}\ \bibnamefont
		{Koivisto}},\ }\href {\doibase 10.1088/1475-7516/2014/02/015} {\bibfield
	{journal} {\bibinfo  {journal} {JCAP}\ }\textbf {\bibinfo {volume} {1402}},\
	\bibinfo {pages} {015} (\bibinfo {year} {2014})},\ \Eprint
{http://arxiv.org/abs/1307.5984} {arXiv:1307.5984 [hep-th]} \BibitemShut
{NoStop}%
%%CITATION = ARXIV:1307.5984;%%
\bibitem [{\citenamefont {Akrami}\ \emph {et~al.}(2014)\citenamefont {Akrami},
	\citenamefont {Koivisto},\ and\ \citenamefont {Solomon}}]{Akrami:2014lja}%
\BibitemOpen
\bibfield  {author} {\bibinfo {author} {\bibfnamefont {Y.}~\bibnamefont
		{Akrami}}, \bibinfo {author} {\bibfnamefont {T.~S.}\ \bibnamefont
		{Koivisto}}, \ and\ \bibinfo {author} {\bibfnamefont {A.~R.}\ \bibnamefont
		{Solomon}},\ }\href {\doibase 10.1007/s10714-014-1838-4} {\bibfield
	{journal} {\bibinfo  {journal} {Gen.Rel.Grav.}\ }\textbf {\bibinfo {volume}
		{47}},\ \bibinfo {pages} {1838} (\bibinfo {year} {2014})},\ \Eprint
{http://arxiv.org/abs/1404.0006} {arXiv:1404.0006 [gr-qc]} \BibitemShut
{NoStop}%
%%CITATION = ARXIV:1404.0006;%%
\bibitem [{\citenamefont {Yamashita}\ \emph {et~al.}(2014)\citenamefont
	{Yamashita}, \citenamefont {De~Felice},\ and\ \citenamefont
	{Tanaka}}]{Yamashita:2014fga}%
\BibitemOpen
\bibfield  {author} {\bibinfo {author} {\bibfnamefont {Y.}~\bibnamefont
		{Yamashita}}, \bibinfo {author} {\bibfnamefont {A.}~\bibnamefont
		{De~Felice}}, \ and\ \bibinfo {author} {\bibfnamefont {T.}~\bibnamefont
		{Tanaka}},\ }\href {\doibase 10.1142/S0218271814430032} {\bibfield  {journal}
	{\bibinfo  {journal} {Int. J. Mod. Phys.}\ }\textbf {\bibinfo {volume}
		{D23}},\ \bibinfo {pages} {1443003} (\bibinfo {year} {2014})},\ \Eprint
{http://arxiv.org/abs/1408.0487} {arXiv:1408.0487 [hep-th]} \BibitemShut
{NoStop}%
%%CITATION = ARXIV:1408.0487;%%
\bibitem [{\citenamefont {Noller}\ and\ \citenamefont
	{Melville}(2015)}]{Noller:2014sta}%
\BibitemOpen
\bibfield  {author} {\bibinfo {author} {\bibfnamefont {J.}~\bibnamefont
		{Noller}}\ and\ \bibinfo {author} {\bibfnamefont {S.}~\bibnamefont
		{Melville}},\ }\href {\doibase 10.1088/1475-7516/2015/01/003} {\bibfield
	{journal} {\bibinfo  {journal} {JCAP}\ }\textbf {\bibinfo {volume} {1501}},\
	\bibinfo {pages} {003} (\bibinfo {year} {2015})},\ \Eprint
{http://arxiv.org/abs/1408.5131} {arXiv:1408.5131 [hep-th]} \BibitemShut
{NoStop}%
%%CITATION = ARXIV:1408.5131;%%
\bibitem [{\citenamefont {de~Rham}\ \emph {et~al.}(2015)\citenamefont
	{de~Rham}, \citenamefont {Heisenberg},\ and\ \citenamefont
	{Ribeiro}}]{deRham:2014naa}%
\BibitemOpen
\bibfield  {author} {\bibinfo {author} {\bibfnamefont {C.}~\bibnamefont
		{de~Rham}}, \bibinfo {author} {\bibfnamefont {L.}~\bibnamefont {Heisenberg}},
	\ and\ \bibinfo {author} {\bibfnamefont {R.~H.}\ \bibnamefont {Ribeiro}},\
}\href {\doibase 10.1088/0264-9381/32/3/035022} {\bibfield  {journal}
{\bibinfo  {journal} {Class. Quant. Grav.}\ }\textbf {\bibinfo {volume}
	{32}},\ \bibinfo {pages} {035022} (\bibinfo {year} {2015})},\ \Eprint
{http://arxiv.org/abs/1408.1678} {arXiv:1408.1678 [hep-th]} \BibitemShut
{NoStop}%
%%CITATION = ARXIV:1408.1678;%%
\bibitem [{\citenamefont {Khosravi}\ \emph {et~al.}(2012)\citenamefont
	{Khosravi}, \citenamefont {Rahmanpour}, \citenamefont {Sepangi},\ and\
	\citenamefont {Shahidi}}]{Khosravi:2011zi}%
\BibitemOpen
\bibfield  {author} {\bibinfo {author} {\bibfnamefont {N.}~\bibnamefont
		{Khosravi}}, \bibinfo {author} {\bibfnamefont {N.}~\bibnamefont
		{Rahmanpour}}, \bibinfo {author} {\bibfnamefont {H.~R.}\ \bibnamefont
		{Sepangi}}, \ and\ \bibinfo {author} {\bibfnamefont {S.}~\bibnamefont
		{Shahidi}},\ }\href {\doibase 10.1103/PhysRevD.85.024049} {\bibfield
	{journal} {\bibinfo  {journal} {Phys.Rev.}\ }\textbf {\bibinfo {volume}
		{D85}},\ \bibinfo {pages} {024049} (\bibinfo {year} {2012})},\ \Eprint
{http://arxiv.org/abs/1111.5346} {arXiv:1111.5346 [hep-th]} \BibitemShut
{NoStop}%
%%CITATION = ARXIV:1111.5346;%%
\bibitem [{\citenamefont {Akrami}\ \emph
	{et~al.}(2013{\natexlab{b}})\citenamefont {Akrami}, \citenamefont
	{Koivisto},\ and\ \citenamefont {Sandstad}}]{Akrami:2012vf}%
\BibitemOpen
\bibfield  {author} {\bibinfo {author} {\bibfnamefont {Y.}~\bibnamefont
		{Akrami}}, \bibinfo {author} {\bibfnamefont {T.~S.}\ \bibnamefont
		{Koivisto}}, \ and\ \bibinfo {author} {\bibfnamefont {M.}~\bibnamefont
		{Sandstad}},\ }\href {\doibase 10.1007/JHEP03(2013)099} {\bibfield  {journal}
	{\bibinfo  {journal} {JHEP}\ }\textbf {\bibinfo {volume} {1303}},\ \bibinfo
	{pages} {099} (\bibinfo {year} {2013}{\natexlab{b}})},\ \Eprint
{http://arxiv.org/abs/1209.0457} {arXiv:1209.0457 [astro-ph.CO]} \BibitemShut
{NoStop}%
%%CITATION = ARXIV:1209.0457;%%
\bibitem [{\citenamefont {Comelli}\ \emph {et~al.}(2014)\citenamefont
	{Comelli}, \citenamefont {Crisostomi},\ and\ \citenamefont
	{Pilo}}]{Comelli:2014bqa}%
\BibitemOpen
\bibfield  {author} {\bibinfo {author} {\bibfnamefont {D.}~\bibnamefont
		{Comelli}}, \bibinfo {author} {\bibfnamefont {M.}~\bibnamefont {Crisostomi}},
	\ and\ \bibinfo {author} {\bibfnamefont {L.}~\bibnamefont {Pilo}},\ }\href
{\doibase 10.1103/PhysRevD.90.084003} {\bibfield  {journal} {\bibinfo
		{journal} {Phys.Rev.}\ }\textbf {\bibinfo {volume} {D90}},\ \bibinfo {pages}
	{084003} (\bibinfo {year} {2014})},\ \Eprint {http://arxiv.org/abs/1403.5679}
{arXiv:1403.5679 [hep-th]} \BibitemShut {NoStop}%
%%CITATION = ARXIV:1403.5679;%%
\bibitem [{\citenamefont {De~Felice}\ \emph {et~al.}(2014)\citenamefont
	{De~Felice}, \citenamefont {Gümrükçüoğlu}, \citenamefont {Mukohyama},
	\citenamefont {Tanahashi},\ and\ \citenamefont {Tanaka}}]{DeFelice:2014nja}%
\BibitemOpen
\bibfield  {author} {\bibinfo {author} {\bibfnamefont {A.}~\bibnamefont
		{De~Felice}}, \bibinfo {author} {\bibfnamefont {A.~E.}\ \bibnamefont
		{Gümrükçüoğlu}}, \bibinfo {author} {\bibfnamefont {S.}~\bibnamefont
		{Mukohyama}}, \bibinfo {author} {\bibfnamefont {N.}~\bibnamefont
		{Tanahashi}}, \ and\ \bibinfo {author} {\bibfnamefont {T.}~\bibnamefont
		{Tanaka}},\ }\href {\doibase 10.1088/1475-7516/2014/06/037} {\bibfield
	{journal} {\bibinfo  {journal} {JCAP}\ }\textbf {\bibinfo {volume} {1406}},\
	\bibinfo {pages} {037} (\bibinfo {year} {2014})},\ \Eprint
{http://arxiv.org/abs/1404.0008} {arXiv:1404.0008 [hep-th]} \BibitemShut
{NoStop}%
%%CITATION = ARXIV:1404.0008;%%
\bibitem [{\citenamefont {Aoki}\ and\ \citenamefont
	{Maeda}(2014)}]{Aoki:2014cla}%
\BibitemOpen
\bibfield  {author} {\bibinfo {author} {\bibfnamefont {K.}~\bibnamefont
		{Aoki}}\ and\ \bibinfo {author} {\bibfnamefont {K.-i.}\ \bibnamefont
		{Maeda}},\ }\href {\doibase 10.1103/PhysRevD.90.124089} {\bibfield  {journal}
	{\bibinfo  {journal} {Phys. Rev.}\ }\textbf {\bibinfo {volume} {D90}},\
	\bibinfo {pages} {124089} (\bibinfo {year} {2014})},\ \Eprint
{http://arxiv.org/abs/1409.0202} {arXiv:1409.0202 [gr-qc]} \BibitemShut
{NoStop}%
%%CITATION = ARXIV:1409.0202;%%
\bibitem [{\citenamefont {Soloviev}(2014)}]{Soloviev:2014eea}%
\BibitemOpen
\bibfield  {author} {\bibinfo {author} {\bibfnamefont {V.~O.}\ \bibnamefont
		{Soloviev}},\ }\href@noop {} {\  (\bibinfo {year} {2014})},\ \Eprint
{http://arxiv.org/abs/1410.0048} {arXiv:1410.0048 [hep-th]} \BibitemShut
{NoStop}%
%%CITATION = ARXIV:1410.0048;%%
\bibitem [{\citenamefont {Heisenberg}(2015)}]{Heisenberg:2014rka}%
\BibitemOpen
\bibfield  {author} {\bibinfo {author} {\bibfnamefont {L.}~\bibnamefont
		{Heisenberg}},\ }\href {\doibase 10.1088/0264-9381/32/10/105011} {\bibfield
	{journal} {\bibinfo  {journal} {Class. Quant. Grav.}\ }\textbf {\bibinfo
		{volume} {32}},\ \bibinfo {pages} {105011} (\bibinfo {year} {2015})},\
\Eprint {http://arxiv.org/abs/1410.4239} {arXiv:1410.4239 [hep-th]}
\BibitemShut {NoStop}%
%%CITATION = ARXIV:1410.4239;%%
\bibitem [{\citenamefont {Hassan}\ \emph {et~al.}(2014)\citenamefont {Hassan},
	\citenamefont {Kocic},\ and\ \citenamefont {Schmidt-May}}]{Hassan:2014gta}%
\BibitemOpen
\bibfield  {author} {\bibinfo {author} {\bibfnamefont {S.}~\bibnamefont
		{Hassan}}, \bibinfo {author} {\bibfnamefont {M.}~\bibnamefont {Kocic}}, \
	and\ \bibinfo {author} {\bibfnamefont {A.}~\bibnamefont {Schmidt-May}},\
}\href@noop {} {\  (\bibinfo {year} {2014})},\ \Eprint
{http://arxiv.org/abs/1409.1909} {arXiv:1409.1909 [hep-th]} \BibitemShut
{NoStop}%
%%CITATION = ARXIV:1409.1909;%%
\bibitem [{\citenamefont {de~Rham}\ \emph
	{et~al.}(2014{\natexlab{b}})\citenamefont {de~Rham}, \citenamefont
	{Heisenberg},\ and\ \citenamefont {Ribeiro}}]{deRham:2014fha}%
\BibitemOpen
\bibfield  {author} {\bibinfo {author} {\bibfnamefont {C.}~\bibnamefont
		{de~Rham}}, \bibinfo {author} {\bibfnamefont {L.}~\bibnamefont {Heisenberg}},
	\ and\ \bibinfo {author} {\bibfnamefont {R.~H.}\ \bibnamefont {Ribeiro}},\
}\href {\doibase 10.1103/PhysRevD.90.124042} {\bibfield  {journal} {\bibinfo
	{journal} {Phys.Rev.}\ }\textbf {\bibinfo {volume} {D90}},\ \bibinfo {pages}
{124042} (\bibinfo {year} {2014}{\natexlab{b}})},\ \Eprint
{http://arxiv.org/abs/1409.3834} {arXiv:1409.3834 [hep-th]} \BibitemShut
{NoStop}%
%%CITATION = ARXIV:1409.3834;%%
\bibitem [{\citenamefont {Enander}\ \emph {et~al.}(2015)\citenamefont
	{Enander}, \citenamefont {Solomon}, \citenamefont {Akrami},\ and\
	\citenamefont {Mortsell}}]{Enander:2014xga}%
\BibitemOpen
\bibfield  {author} {\bibinfo {author} {\bibfnamefont {J.}~\bibnamefont
		{Enander}}, \bibinfo {author} {\bibfnamefont {A.~R.}\ \bibnamefont
		{Solomon}}, \bibinfo {author} {\bibfnamefont {Y.}~\bibnamefont {Akrami}}, \
	and\ \bibinfo {author} {\bibfnamefont {E.}~\bibnamefont {Mortsell}},\ }\href
{\doibase 10.1088/1475-7516/2015/01/006} {\bibfield  {journal} {\bibinfo
		{journal} {JCAP}\ }\textbf {\bibinfo {volume} {1501}},\ \bibinfo {pages}
	{006} (\bibinfo {year} {2015})},\ \Eprint {http://arxiv.org/abs/1409.2860}
{arXiv:1409.2860 [astro-ph.CO]} \BibitemShut {NoStop}%
%%CITATION = ARXIV:1409.2860;%%
\bibitem [{\citenamefont {Schmidt-May}(2015)}]{Schmidt-May:2014xla}%
\BibitemOpen
\bibfield  {author} {\bibinfo {author} {\bibfnamefont {A.}~\bibnamefont
		{Schmidt-May}},\ }\href {\doibase 10.1088/1475-7516/2015/01/039} {\bibfield
	{journal} {\bibinfo  {journal} {JCAP}\ }\textbf {\bibinfo {volume} {1501}},\
	\bibinfo {pages} {039} (\bibinfo {year} {2015})},\ \Eprint
{http://arxiv.org/abs/1409.3146} {arXiv:1409.3146 [gr-qc]} \BibitemShut
{NoStop}%
%%CITATION = ARXIV:1409.3146;%%
\bibitem [{\citenamefont {Emir~Gümrükçüoğlu}\ \emph
	{et~al.}(2015)\citenamefont {Emir~Gümrükçüoğlu}, \citenamefont
	{Heisenberg},\ and\ \citenamefont {Mukohyama}}]{Gumrukcuoglu:2014xba}%
\BibitemOpen
\bibfield  {author} {\bibinfo {author} {\bibfnamefont {A.}~\bibnamefont
		{Emir~Gümrükçüoğlu}}, \bibinfo {author} {\bibfnamefont {L.}~\bibnamefont
		{Heisenberg}}, \ and\ \bibinfo {author} {\bibfnamefont {S.}~\bibnamefont
		{Mukohyama}},\ }\href {\doibase 10.1088/1475-7516/2015/02/022} {\bibfield
	{journal} {\bibinfo  {journal} {JCAP}\ }\textbf {\bibinfo {volume} {1502}},\
	\bibinfo {pages} {022} (\bibinfo {year} {2015})},\ \Eprint
{http://arxiv.org/abs/1409.7260} {arXiv:1409.7260 [hep-th]} \BibitemShut
{NoStop}%
%%CITATION = ARXIV:1409.7260;%%
\bibitem [{\citenamefont {Noller}(2015)}]{Noller:2014ioa}%
\BibitemOpen
\bibfield  {author} {\bibinfo {author} {\bibfnamefont {J.}~\bibnamefont
		{Noller}},\ }\href {\doibase 10.1088/1475-7516/2015/04/025} {\bibfield
	{journal} {\bibinfo  {journal} {JCAP}\ }\textbf {\bibinfo {volume} {1504}},\
	\bibinfo {pages} {025} (\bibinfo {year} {2015})},\ \Eprint
{http://arxiv.org/abs/1409.7692} {arXiv:1409.7692 [hep-th]} \BibitemShut
{NoStop}%
%%CITATION = ARXIV:1409.7692;%%
\bibitem [{\citenamefont {Solomon}\ \emph {et~al.}(2015)\citenamefont
	{Solomon}, \citenamefont {Enander}, \citenamefont {Akrami}, \citenamefont
	{Koivisto}, \citenamefont {Könnig},\ and\ \citenamefont
	{Mörtsell}}]{Solomon:2014iwa}%
\BibitemOpen
\bibfield  {author} {\bibinfo {author} {\bibfnamefont {A.~R.}\ \bibnamefont
		{Solomon}}, \bibinfo {author} {\bibfnamefont {J.}~\bibnamefont {Enander}},
	\bibinfo {author} {\bibfnamefont {Y.}~\bibnamefont {Akrami}}, \bibinfo
	{author} {\bibfnamefont {T.~S.}\ \bibnamefont {Koivisto}}, \bibinfo {author}
	{\bibfnamefont {F.}~\bibnamefont {Könnig}}, \ and\ \bibinfo {author}
	{\bibfnamefont {E.}~\bibnamefont {Mörtsell}},\ }\href {\doibase
	10.1088/1475-7516/2015/04/027} {\bibfield  {journal} {\bibinfo  {journal}
		{JCAP}\ }\textbf {\bibinfo {volume} {1504}},\ \bibinfo {pages} {027}
	(\bibinfo {year} {2015})},\ \Eprint {http://arxiv.org/abs/1409.8300}
{arXiv:1409.8300 [astro-ph.CO]} \BibitemShut {NoStop}%
%%CITATION = ARXIV:1409.8300;%%
\bibitem [{\citenamefont {Mukohyama}(2014)}]{Mukohyama:2014rca}%
\BibitemOpen
\bibfield  {author} {\bibinfo {author} {\bibfnamefont {S.}~\bibnamefont
		{Mukohyama}},\ }\href {\doibase 10.1088/1475-7516/2014/12/011} {\bibfield
	{journal} {\bibinfo  {journal} {JCAP}\ }\textbf {\bibinfo {volume} {1412}},\
	\bibinfo {pages} {011} (\bibinfo {year} {2014})},\ \Eprint
{http://arxiv.org/abs/1410.1996} {arXiv:1410.1996 [hep-th]} \BibitemShut
{NoStop}%
%%CITATION = ARXIV:1410.1996;%%
\bibitem [{\citenamefont {Gabadadze}\ \emph {et~al.}(2012)\citenamefont
	{Gabadadze}, \citenamefont {Hinterbichler}, \citenamefont {Khoury},
	\citenamefont {Pirtskhalava},\ and\ \citenamefont
	{Trodden}}]{Gabadadze:2012tr}%
\BibitemOpen
\bibfield  {author} {\bibinfo {author} {\bibfnamefont {G.}~\bibnamefont
		{Gabadadze}}, \bibinfo {author} {\bibfnamefont {K.}~\bibnamefont
		{Hinterbichler}}, \bibinfo {author} {\bibfnamefont {J.}~\bibnamefont
		{Khoury}}, \bibinfo {author} {\bibfnamefont {D.}~\bibnamefont
		{Pirtskhalava}}, \ and\ \bibinfo {author} {\bibfnamefont {M.}~\bibnamefont
		{Trodden}},\ }\href {\doibase 10.1103/PhysRevD.86.124004} {\bibfield
	{journal} {\bibinfo  {journal} {Phys.Rev.}\ }\textbf {\bibinfo {volume}
		{D86}},\ \bibinfo {pages} {124004} (\bibinfo {year} {2012})},\ \Eprint
{http://arxiv.org/abs/1208.5773} {arXiv:1208.5773 [hep-th]} \BibitemShut
{NoStop}%
%%CITATION = ARXIV:1208.5773;%%
\bibitem [{\citenamefont {Andrews}\ \emph
	{et~al.}(2013{\natexlab{a}})\citenamefont {Andrews}, \citenamefont {Goon},
	\citenamefont {Hinterbichler}, \citenamefont {Stokes},\ and\ \citenamefont
	{Trodden}}]{Andrews:2013ora}%
\BibitemOpen
\bibfield  {author} {\bibinfo {author} {\bibfnamefont {M.}~\bibnamefont
		{Andrews}}, \bibinfo {author} {\bibfnamefont {G.}~\bibnamefont {Goon}},
	\bibinfo {author} {\bibfnamefont {K.}~\bibnamefont {Hinterbichler}}, \bibinfo
	{author} {\bibfnamefont {J.}~\bibnamefont {Stokes}}, \ and\ \bibinfo {author}
	{\bibfnamefont {M.}~\bibnamefont {Trodden}},\ }\href {\doibase
	10.1103/PhysRevLett.111.061107} {\bibfield  {journal} {\bibinfo  {journal}
		{Phys.Rev.Lett.}\ }\textbf {\bibinfo {volume} {111}},\ \bibinfo {pages}
	{061107} (\bibinfo {year} {2013}{\natexlab{a}})},\ \Eprint
{http://arxiv.org/abs/1303.1177} {arXiv:1303.1177 [hep-th]} \BibitemShut
{NoStop}%
%%CITATION = ARXIV:1303.1177;%%
\bibitem [{\citenamefont {Andrews}\ \emph
	{et~al.}(2013{\natexlab{b}})\citenamefont {Andrews}, \citenamefont
	{Hinterbichler}, \citenamefont {Stokes},\ and\ \citenamefont
	{Trodden}}]{Andrews:2013uca}%
\BibitemOpen
\bibfield  {author} {\bibinfo {author} {\bibfnamefont {M.}~\bibnamefont
		{Andrews}}, \bibinfo {author} {\bibfnamefont {K.}~\bibnamefont
		{Hinterbichler}}, \bibinfo {author} {\bibfnamefont {J.}~\bibnamefont
		{Stokes}}, \ and\ \bibinfo {author} {\bibfnamefont {M.}~\bibnamefont
		{Trodden}},\ }\href {\doibase 10.1088/0264-9381/30/18/184006} {\bibfield
	{journal} {\bibinfo  {journal} {Class.Quant.Grav.}\ }\textbf {\bibinfo
		{volume} {30}},\ \bibinfo {pages} {184006} (\bibinfo {year}
	{2013}{\natexlab{b}})},\ \Eprint {http://arxiv.org/abs/1306.5743}
{arXiv:1306.5743 [hep-th]} \BibitemShut {NoStop}%
%%CITATION = ARXIV:1306.5743;%%
\bibitem [{\citenamefont {Goon}\ \emph {et~al.}(2014)\citenamefont {Goon},
	\citenamefont {Gümrükçüoğlu}, \citenamefont {Hinterbichler},
	\citenamefont {Mukohyama},\ and\ \citenamefont {Trodden}}]{Goon:2014ywa}%
\BibitemOpen
\bibfield  {author} {\bibinfo {author} {\bibfnamefont {G.}~\bibnamefont
		{Goon}}, \bibinfo {author} {\bibfnamefont {A.~E.}\ \bibnamefont
		{Gümrükçüoğlu}}, \bibinfo {author} {\bibfnamefont {K.}~\bibnamefont
		{Hinterbichler}}, \bibinfo {author} {\bibfnamefont {S.}~\bibnamefont
		{Mukohyama}}, \ and\ \bibinfo {author} {\bibfnamefont {M.}~\bibnamefont
		{Trodden}},\ }\href {\doibase 10.1088/1475-7516/2014/08/008} {\bibfield
	{journal} {\bibinfo  {journal} {JCAP}\ }\textbf {\bibinfo {volume} {1408}},\
	\bibinfo {pages} {008} (\bibinfo {year} {2014})},\ \Eprint
{http://arxiv.org/abs/1402.5424} {arXiv:1402.5424 [hep-th]} \BibitemShut
{NoStop}%
%%CITATION = ARXIV:1402.5424;%%
\bibitem [{\citenamefont {Gao}\ and\ \citenamefont {Steer}(2011)}]{Gao:2011qe}%
\BibitemOpen
\bibfield  {author} {\bibinfo {author} {\bibfnamefont {X.}~\bibnamefont
		{Gao}}\ and\ \bibinfo {author} {\bibfnamefont {D.~A.}\ \bibnamefont
		{Steer}},\ }\href {\doibase 10.1088/1475-7516/2011/12/019} {\bibfield
	{journal} {\bibinfo  {journal} {JCAP}\ }\textbf {\bibinfo {volume} {1112}},\
	\bibinfo {pages} {019} (\bibinfo {year} {2011})},\ \Eprint
{http://arxiv.org/abs/1107.2642} {arXiv:1107.2642 [astro-ph.CO]} \BibitemShut
{NoStop}%
%%CITATION = ARXIV:1107.2642;%%
\bibitem [{\citenamefont {Gao}(2011)}]{Gao:2011mz}%
\BibitemOpen
\bibfield  {author} {\bibinfo {author} {\bibfnamefont {X.}~\bibnamefont
		{Gao}},\ }\href {\doibase 10.1088/1475-7516/2011/10/021} {\bibfield
	{journal} {\bibinfo  {journal} {JCAP}\ }\textbf {\bibinfo {volume} {1110}},\
	\bibinfo {pages} {021} (\bibinfo {year} {2011})},\ \Eprint
{http://arxiv.org/abs/1106.0292} {arXiv:1106.0292 [astro-ph.CO]} \BibitemShut
{NoStop}%
%%CITATION = ARXIV:1106.0292;%%
\bibitem [{\citenamefont {De~Felice}\ and\ \citenamefont
	{Tsujikawa}(2011)}]{DeFelice:2011uc}%
\BibitemOpen
\bibfield  {author} {\bibinfo {author} {\bibfnamefont {A.}~\bibnamefont
		{De~Felice}}\ and\ \bibinfo {author} {\bibfnamefont {S.}~\bibnamefont
		{Tsujikawa}},\ }\href {\doibase 10.1103/PhysRevD.84.083504} {\bibfield
	{journal} {\bibinfo  {journal} {Phys.Rev.}\ }\textbf {\bibinfo {volume}
		{D84}},\ \bibinfo {pages} {083504} (\bibinfo {year} {2011})},\ \Eprint
{http://arxiv.org/abs/1107.3917} {arXiv:1107.3917 [gr-qc]} \BibitemShut
{NoStop}%
%%CITATION = ARXIV:1107.3917;%%
\bibitem [{\citenamefont {Folkerts}\ \emph {et~al.}(2011)\citenamefont
	{Folkerts}, \citenamefont {Pritzel},\ and\ \citenamefont
	{Wintergerst}}]{Folkerts:2011ev}%
\BibitemOpen
\bibfield  {author} {\bibinfo {author} {\bibfnamefont {S.}~\bibnamefont
		{Folkerts}}, \bibinfo {author} {\bibfnamefont {A.}~\bibnamefont {Pritzel}}, \
	and\ \bibinfo {author} {\bibfnamefont {N.}~\bibnamefont {Wintergerst}},\
}\href@noop {} {\  (\bibinfo {year} {2011})},\ \Eprint
{http://arxiv.org/abs/1107.3157} {arXiv:1107.3157 [hep-th]} \BibitemShut
{NoStop}%
%%CITATION = ARXIV:1107.3157;%%
\bibitem [{\citenamefont {Hinterbichler}(2013)}]{Hinterbichler:2013eza}%
\BibitemOpen
\bibfield  {author} {\bibinfo {author} {\bibfnamefont {K.}~\bibnamefont
		{Hinterbichler}},\ }\href {\doibase 10.1007/JHEP10(2013)102} {\bibfield
	{journal} {\bibinfo  {journal} {JHEP}\ }\textbf {\bibinfo {volume} {1310}},\
	\bibinfo {pages} {102} (\bibinfo {year} {2013})},\ \Eprint
{http://arxiv.org/abs/1305.7227} {arXiv:1305.7227 [hep-th]} \BibitemShut
{NoStop}%
%%CITATION = ARXIV:1305.7227;%%
\bibitem [{\citenamefont {Kimura}\ and\ \citenamefont
	{Yamauchi}(2013)}]{Kimura:2013ika}%
\BibitemOpen
\bibfield  {author} {\bibinfo {author} {\bibfnamefont {R.}~\bibnamefont
		{Kimura}}\ and\ \bibinfo {author} {\bibfnamefont {D.}~\bibnamefont
		{Yamauchi}},\ }\href {\doibase 10.1103/PhysRevD.88.084025} {\bibfield
	{journal} {\bibinfo  {journal} {Phys.Rev.}\ }\textbf {\bibinfo {volume}
		{D88}},\ \bibinfo {pages} {084025} (\bibinfo {year} {2013})},\ \Eprint
{http://arxiv.org/abs/1308.0523} {arXiv:1308.0523 [gr-qc]} \BibitemShut
{NoStop}%
%%CITATION = ARXIV:1308.0523;%%
\bibitem [{\citenamefont {de~Rham}\ \emph
	{et~al.}(2014{\natexlab{c}})\citenamefont {de~Rham}, \citenamefont {Matas},\
	and\ \citenamefont {Tolley}}]{deRham:2013tfa}%
\BibitemOpen
\bibfield  {author} {\bibinfo {author} {\bibfnamefont {C.}~\bibnamefont
		{de~Rham}}, \bibinfo {author} {\bibfnamefont {A.}~\bibnamefont {Matas}}, \
	and\ \bibinfo {author} {\bibfnamefont {A.~J.}\ \bibnamefont {Tolley}},\
}\href {\doibase 10.1088/0264-9381/31/16/165004} {\bibfield  {journal}
{\bibinfo  {journal} {Class.Quant.Grav.}\ }\textbf {\bibinfo {volume} {31}},\
\bibinfo {pages} {165004} (\bibinfo {year} {2014}{\natexlab{c}})},\ \Eprint
{http://arxiv.org/abs/1311.6485} {arXiv:1311.6485 [hep-th]} \BibitemShut
{NoStop}%
%%CITATION = ARXIV:1311.6485;%%
\bibitem [{\citenamefont {Gao}(2014)}]{Gao:2014jja}%
\BibitemOpen
\bibfield  {author} {\bibinfo {author} {\bibfnamefont {X.}~\bibnamefont
		{Gao}},\ }\href {\doibase 10.1103/PhysRevD.90.064024} {\bibfield  {journal}
	{\bibinfo  {journal} {Phys.Rev.}\ }\textbf {\bibinfo {volume} {D90}},\
	\bibinfo {pages} {064024} (\bibinfo {year} {2014})},\ \Eprint
{http://arxiv.org/abs/1403.6781} {arXiv:1403.6781 [hep-th]} \BibitemShut
{NoStop}%
%%CITATION = ARXIV:1403.6781;%%
\end{thebibliography}
\end{document}